\documentclass[aps,prd,preprint,nofootinbib]{revtex4}
\usepackage{epsfig}
\usepackage{latexsym,amsmath,amssymb,amstext}
\usepackage{float}
\usepackage{soul}
\usepackage{slashed}
\usepackage[paperwidth=210mm,paperheight=297mm,centering,hmargin=2cm,vmargin=2.5cm]{geometry}

\newcommand{\be}{\begin{equation}}
\newcommand{\ee}{\end{equation}}
\newcommand{\bea}{\begin{eqnarray}}
\newcommand{\eea}{\end{eqnarray}}
\newcommand{\Tr}{{\rm Tr}}

\newcommand\bef{\begin{figure}}
\newcommand\eef[1]{\label{fg:#1}\end{figure}}
\newcommand\beq{\begin{equation}}
\newcommand\eeq[1]{\label{#1}\end{equation}}
\newcommand\beqa{\begin{eqnarray}}
\newcommand\eeqa[1]{\label{#1}\end{eqnarray}}
\newcommand\bet{\begin{table}}
\newcommand\eet[1]{\label{tb:#1}\end{table}}

\newcommand\fgn[1]{Figure \ref{fg:#1}}
\newcommand\eqn[1]{Eq.\ (\ref{#1})}

\newcommand\apx[1]{Appendix \ref{sec:#1}}

\begin{document}

\date{\today}

\title{
Phase diagram of the large $N$ Gross-Neveu model in  a finite periodic box
}
\author{R. Narayanan}
\email{rajamani.narayanan@fiu.edu}
\affiliation{Department of Physics, Florida International University, Miami,
FL 33199.}
\begin{abstract}
We analyze the Gross-Neveu model in the limit of large number of flavors of massless fermions.
We study the phase diagram  in a two and three dimensional periodic box at a fixed thermal to spatial aspect ratio, $\frac{\beta}{\ell}$,  with
a flavor independent chemical potential.
We assume the bilinear condensate, when one exists, has a specific momentum in the spatial direction(s). The main known features of the phase
diagram 
 in the $\ell\to\infty$ limit  of the two dimensional model are also seen on a finite $\ell\times\beta$ torus --  a phase with a
homogeneous (zero momentum) condensate; a phase with an inhomogeneous (non-zero momentum) condensate and a phase with no condensate. We observe that the inhomogeneous phase 
contains several sub-phases characterized by a specific spatial momentum.  Unlike the two dimensional model, we do not find evidence for
a phase with a inhomogeneous condensate in the three dimensional model.
\end{abstract}

\maketitle

\section{Introduction}

The two dimensional Gross-Neveu model~\cite{Gross:1974jv} with $N$ flavors of massless fermions and a flavor-singlet quartic self-interaction has been shown to be quite an interesting toy model
with several features that mimic four dimensional QCD. Of particular interest for us is the phase diagram of this model in the $\mu-T$ (chemical potential - temperature) plane in the $N\to\infty$ limit. The original analysis~\cite{Wolff:1985av} showed a line of transitions that starts out as a second-order transition at $(0,T_c)$ and ends up as a first-order transition at $(\mu_c,0)$ with a tri-critical point at $(\mu_T,T_T)$ that separates the
first-order line from the second-order line. Later on, a complete Hartree-Fock calculation showed the presence of a phase with an inhomogeneous condensate~\cite{Thies:2003kk}. This result was confirmed later using a lattice computation~\cite{deForcrand:2006zz} that performed a brute force minimization of the
free energy and studied instabilities. More recently, this model was numerically studied on the lattice at finite but large number of flavors (eight to be exact)~\cite{Pannullo:2019prx} and was shown to have a phase with an inhomogeneous condensate.  

The three dimensional Gross-Neveu model was shown to be renormalizable in a $\frac{1}{N}$ expansion~\cite{Gross:1975vu,Parisi:1975im,Shizuya:1979bv,Rosenstein:1988pt} and the phase diagram in the
$N\to\infty$ limit has a a line of second order transitions~\cite{Klimenko:1987gi,Rosenstein:1988dj}.
Possibility of a phase with an inhomogeneous condensate has been recently explored with negative~\cite{Urlichs:2007zz} and positive~\cite{Winstel:2019zfn} results.

Lattice formalism is one of the best ways to find a global minimum when the action density has many local minimums.
Motivated by the clear observation in~\cite{Pannullo:2019prx} that the condensate in the inhomogeneous phase
has a specific momentum and amplitude that varies with the value of the chemical potential and temperature, we study the effective action
of the condensate at a fixed momentum in the limit of large number of flavors. We use the lattice regularization to compute the fermion determinant and study the continuum limit in a  periodic box with a spatial extent $\ell$ and thermal extent $\beta$. We will show that the inhomogeneous phase in the two dimensional
model at a fixed aspect ratio, $\tau=\frac{\beta}{\ell}$, contains several sub-phases characterized by a specific momentum that 
depends on the temperature and chemical potential.
The results in three dimensions are different. We will show that a region in the $(\mu, T)$ plane sustains a inhomogeneous condensate
at a finite lattice spacing consistent with the observation in~\cite{Winstel:2019zfn} but this region shrinks with decreasing lattice spacing suggesting that an inhomogeneous phase does not exist
in the continuum limit supporting the analysis in~\cite{Urlichs:2007zz}. 

\section{Action density on the lattice at a fixed momentum}

Our aim in this section is to obtain the effective action for the condensate at a fixed momentum within the lattice regularization.
The main technical point is the use of a formula for the determinant of a block tri-diagonal matrix~\cite{2007arXiv0712.0681M}.

The action for the continuum Gross-Neveu model on a periodic box with a finite chemical potential is given by
\be
S(\bar\psi_i,\psi_i;\lambda) = \int [dx] \left[ \sum_{i=1}^N \bar \psi_i D(\mu)  \psi _i - \frac{1}{2N\lambda}\left( \sum_{i=1}^N \bar\psi_i \psi_i\right)^2 \right]
,\label{gn2d}
\ee
where the massless Dirac operator is 
\be
D(\mu) =\begin{cases}
 \sigma_1 \partial_1 + \sigma_2 \left(\partial_2 + \mu\right) & {\rm on\ } \ell\times\beta ,\cr
\sigma_1 \partial_1 + \sigma_2\partial_2 + \sigma_3 \left(\partial_3 + \mu\right) & {\rm on\ } \ell^2\times\beta.
\end{cases}
\ee
 The fermion fields obey periodic boundary conditions in the spatial direction(s) and anti-periodic boundary conditions in the
thermal direction.
Upon introduction of a scalar auxiliary field, $M(x)$, to replace the four-fermi interaction, and a subsequent integration of the fermions
results in 
\be
S(M;\lambda) = N \int d^2 x \left[ \frac{\lambda}{2} M^2 - \ln \det  (M+D(\mu)) \right],\label{gnaction}
\ee
as the effective action for the scalar field.

Our aim in this paper is to study the minimum of the action with the scalar field restricted to
\be
M(x) =\begin{cases}
m^L_q \cos \frac{2 \pi qx_1}{\ell} & {\rm on\ } \ell\times\beta ,\cr
m^L_{\bf q} \cos \frac{2 \pi (q_1x_1+q_2x_2)}{\ell} & {\rm on\ } \ell^2\times\beta .
\end{cases}
\label{sigmom}
\ee
We will refer to the integer $q$ as the wavenumber and the pair of integers $(q_1,q_2)$ as the wavevector~\footnote{This is different from the standard convention by a factor of $\ell$.}.
We will find the value of the momentum (wavenumber) at which the action has a minimum taken over a wide range of momenta (wavenumber).
Since we are studying the theory in the large $N$ limit, lattice doublers are not relevant and 
we will use the lattice regularization with naive fermions to compute the fermion determinant with $L$ points in the spatial
direction(s) and $L_T$ points in the thermal direction. We will restrict ourselves to prime numbers for $L$ and even values for $L_T$.
This restriction will become evident when we describe the technical details.
We will use different aspect ratios
\be
\tau=
\frac{\beta}{\ell} = \frac{1}{\ell T} = \frac{L_T}{L}
\ee
to study the phase diagram. We will analyze the two dimensional model
for seven different aspect ratios (three of these will be used to analyze the three dimensional model), namely,
\be
\frac{\beta}{\ell} = \left ( \frac{40}{127}, \frac{50}{127}, \frac{60}{127}, \frac{70}{127}, \frac{80}{127}, \frac{90}{127}, \frac{100}{127} \right).
\label{taulist}
\ee
Four different sets of $(L,L_T)$ at each aspect ratio given by
\bea
\tau=\frac{40}{127} &:& (127,40), (151,48), (179,56), (197,62) \cr
\tau=\frac{50}{127} &:& (127,50), (157,62), (179,70), (199,78) \cr
\tau=\frac{60}{127} &:& (127,60), (149,70), (173,82), (199,94) \cr
\tau=\frac{70}{127} &:& (127,70), (149,82), (167,92), (199,110) \cr
\tau=\frac{80}{127} &:& (127,80), (149,94), (181,114), (193,122) \cr
\tau=\frac{90}{127} &:& (127,90), (149,106), (167,118), (197,140) \cr
\tau=\frac{100}{127} &:& (127,100), (157,124), (173,136), (197,152) ,\label{lltlist}
\eea
will be used to study finite lattice spacing effects at a fixed aspect ratio. The phase diagram is usually studied at infinite spatial extent
and this corresponds to a zero aspect ratio  at non-zero temperature. Therefore, we will study effects due to non-zero aspect ratio
after we have taken the continuum limit at a fixed aspect ratio.

 The massless naive lattice Dirac operator
in momentum space with the the chemical potential
on the lattice, $\mu_L$, 
introduced as a constant imaginary vector potential in the thermal direction~\cite{Hasenfratz:1983ba,Kogut:1983ia} is
\be
D({\bf k},\mu_L) = \begin{cases}
\begin{pmatrix} 0 & c({\bf k},\mu_L) \cr -c^*({\bf k},-\mu_L) & 0 \end{pmatrix} 
& {\rm on\ } L\times L_T ,\cr
i\sigma_1 \sin \frac{2\pi k_1}{L} +i \sigma_2 \sin\frac{2\pi k_2}{L} + i\sigma_3\sin \left[ \frac{\pi(2k_3+1)}{L_T} - i\mu_L\right] & {\rm on\ } L^2\times L_T ,
\end{cases}
\ee
where
\be
c({\bf k},\mu_L) = i \sin \frac{2\pi k_1}{L} + \sin \left[ \frac{\pi(2k_2+1)}{L_T} - i\mu_L\right],
\ee
is the chiral Dirac operator in two dimensions.

The  action density for $m^L_0$ with $q=0$ is
\be
s_0(m^L_0;\lambda,\mu_L,L,L_T) = 
\begin{cases}
\frac{1}{2}\lambda {\left(m^L_0\right)}^2 - \frac{1}{LL_T} \sum_{\bf k} \ln [ d_2({\bf k},\mu_L) +{\left(m^L_0\right)}^2],
& {\rm on\ } L\times L_T ,\cr
\frac{1}{2}\lambda {\left(m^L_0\right)}^2 - \frac{1}{L^2L_T} \sum_{\bf k} \ln [ d_3({\bf k},\mu_L)+{\left(m^L_0\right)}^2],
& {\rm on\ } L^2\times L_T .\cr
\end{cases}
\label{actden}
\ee
where
\bea
d_2({\bf k},\mu_L) &=& \sin^2 \frac{2\pi k_1}{L} +\sin^2 \left[ \frac{\pi(2k_2+1)}{L_T} - i\mu_L\right] ;\cr
d_3({\bf k},\mu_L) &=& \sin^2 \frac{2\pi k_1}{L} +\sin^2 \frac{2\pi k_2}{L} + \sin^2 \left[ \frac{\pi(2k_3+1)}{L_T} - i\mu_L\right] .
\eea
The fermion determinant is real and positive in two and three dimensions
since we have a pair of momenta with $k_2+k'_2=-1$ in two dimensions and a pair of momenta with $k_3+k'_3=-1$ in three dimensions.

When ${\bf q}$ is non-zero in \eqn{sigmom}, the fermion determinant still factorizes into momenta in the thermal direction but it couples
$k_1$ with $k_1+q$ in the spatial direction of the two dimensional model and couples $(k_1,k_2)$ with $(k_1+q_1,k_2+q_2)$ in the
spatial directions of the three dimensional model.  We define a sequence of momenta for the two dimensional model by $k_1^j = j q$ for $j=0,1,\cdots$. This sequence will always be a cycle of length $L$   and couple all momenta in the spatial direction
since the cycle will terminate only when $p q$ is a multiple of $L$ for some integer $p$. Since we have assumed that $L$ is prime and $q \in [0,L-1]$, we conclude that $p=L$. By a similar argument, a sequence of momenta for the three dimensional model defined by
$(k_1^j,k_2^j) = (jq_1,k_2+jq_2)$ for each choice of $k_2\in [0,L-1]$ will be of length $L$ as long as $q_1\ne 0$ which we assume to be the
case. The sets of $L$ momenta for different choices of $k_2$ will be disjoint. For brevity, we define
\be
D_j = \begin{cases} D(k_1^j,k_2,\mu_L) & {\rm on\ } L\times L_T ,\cr
D(k_1^j,k_2^j,k_3,\mu_L) & {\rm on\ } L^2\times L_T.
\end{cases}
\ee
The fermion determinant for each cycle of length $L$ is
\be
\left.
\begin{aligned}
\det D_c (k_2,m^L_q,\mu_L)  \quad {\rm on\ } L\times L_T \cr
\det D_c (k_2,k_3,m^L_q,\mu_L)  \quad {\rm on\ } L^2\times L_T 
\end{aligned}
\right\}
= \det \begin{pmatrix}
D_0 & \frac{1}{2}m^L_q & 0 & \cdots & 0 & 0 & \frac{1}{2}m^L_q \cr
\frac{1}{2}m^L_q & D_1  & \frac{1}{2}m^L_q & \cdots & 0 & 0 & 0 \cr
0 & \frac{1}{2}m^L_q & D_2 & \cdots & 0 & 0 & 0 \cr
\vdots & \vdots & \vdots & \ddots & \vdots & \vdots & \vdots \cr
0 & 0 & 0 & \cdots & D_{L-3} & \frac{1}{2}m^L_q & 0\cr
0 & 0 & 0 & \cdots & \frac{1}{2}m^L_q & D_{L-2} & \frac{1}{2}m^L_q \cr
\frac{1}{2}m^L_q & 0 & 0 & \cdots & 0 & \frac{1}{2}m^L_q & D_{L-1} 
\end{pmatrix}.\label{fermat}
\ee
The action density for $m^L_q$ with $q\ne 0$ is
\be
s_q(m^L_q;\lambda,\mu_L,L,L_T) = 
\begin{cases}
\frac{1}{4}\lambda {\left(m^L_q\right)}^2 - \frac{1}{LL_T} \sum_{k_2} \ln [\det D_c (k_2,m^L_q,\mu_L)],
& {\rm on\ } L\times L_T ,\cr
\frac{1}{4}\lambda {\left(m^L_q\right)}^2 - \frac{1}{L^2L_T} \sum_{k_2,k_3} \ln [\det D_c (k_2,k_3,m^L_q,\mu_L) ],
& {\rm on\ } L^2\times L_T .\cr
\end{cases}.
\label{actdenq}
\ee
Like in the $q=0$ case, the fermion determinant after appropriating pairing of momenta is real and positive.

We need to rewrite the formula for the determinant obtained from~\cite{2007arXiv0712.0681M} such that only positive powers of
$m^L_q$ appear in the result. The expression for the three dimensional model and a further simplification fo the two dimensional
model due to the chiral structure of the massless Dirac operator are derived in \apx{detqform}. The final result for the three dimensional
model is
\be
\det D_c(k_2,k_3,\mu_L) = - \left[ 1-  \left (  \frac{m^L_q}{2} \right)^{2L}\right]^2 - \left ( - \frac{m^L_q}{2} \right)^{L} \Tr \left( \bar T +  \tilde T \right)
+ \frac{1}{2} \left [ \det (\bar T -1) + \det (\bar T+1) \right] ,
\ee
where
\bea
\bar T =\bar T_{L-1} \bar T_{L-2} \cdots \bar T_1 \bar T_0;
\quad && \quad
\tilde T = \tilde T_0 \tilde T_1 \cdots \tilde T_{L-2} \tilde T_{L-1}\cr
\bar T_i = \begin{pmatrix} D_i & -\frac{1}{4} {m^L_q}^2 \cr 1 & 0 \end{pmatrix}; \quad && \quad
\tilde T_i = \begin{pmatrix} 0 & \frac{1}{4} {m^L_q}^2 \cr -1 &  D_i \end{pmatrix}.
\eea
The final result for the two dimensional model is
\be
\det D_c(k_2,\mu_L) = \left [ 2\left( \frac{\Sigma}{2} \right) ^{2L} - \Tr (\Psi\Phi) \right],
\ee
where
\bea
\Psi &=& X_{L-1} Y_{L-2} X_{L-3} Y_{L-4} \cdots X_2 Y_1 X_0;\cr
\Phi  &=& Y_{L-1} X_{L-2} Y_{L-3} X_{L-4} \cdots Y_2 X_1 Y_0.
\eea
\be
X_i = \begin{pmatrix} c(k_1^i,k_2,\mu_L) & -\frac{\Sigma^2}{4} \cr 1 & 0 \end{pmatrix};\qquad Y_i = \begin{pmatrix} -c^*(k_1^i,k_2,-\mu_L) & -\frac{\Sigma^2}{4} \cr 1 & 0 \end{pmatrix}.
\ee

\section{Results}

The gap equation at zero momentum and zero chemical potential with $L=L_T$ given by
\be
\lambda(m_s) = \begin{cases}
\frac{2}{L^2} \sum_{\bf k} \left(d_2({\bf k},0) +m_s^2\right)^{-1} & {\rm on\ } L^2 \cr
\frac{2}{L^3} \sum_{\bf k} \left(d_3({\bf k},0) +m_s^2\right)^{-1} & {\rm on\ } L^3
\end{cases}\label{copscale}
\ee
can be used to map the lattice coupling, $\lambda$, to a lattice scale, $m_s$. A subtle point to note is that $\lambda(0)$ approaches
infinity as $\ln L$ on $L^2$ like in a continuum regularization but it approaches a finite value on $L^3$ unlike in a continuum regularization
where we will see a linear divergence. Having chosen an aspect ratio $\tau$, we fix a pair $(L,L_T)$ that obeys the aspect ratio.
Next, we pick a temperature, $T$, and a chemical potential, $\mu$ that are dimensionless -- measured in units of the scale.
The lattice scale is given by
\be
m_s = \frac{1}{L_T T},
\ee
and the chemical potential on the lattice is given by
\be
\mu_L = \mu m_s.
\ee
We use \eqn{copscale} to find $\lambda(m_s)$. We find the minimum of the action density, $s_q(m_qm_s;\lambda(m_s),\mu m_s, L, L_T)$,
as a function of $m_q$
at a fixed $q$ and find the value of $q$ where the minimum value is the lowest. We look at $q\in [0,10]$ in two dimensions to find
the minimum. We look at 
\be
(q_1,q_2)=(0,0),(1,0),(1,1),(2,0),(2,1),(2,2),(3,0),(3,1),(3,2),(4,0),(4,1)\label{momlist3d}
\ee
in order of increasing magnitude in three dimensions. This range of momenta was sufficient to obtain a picture of the phase diagram.

\subsection{Two dimensional model}

\bef
\centering
\includegraphics[scale=0.325]{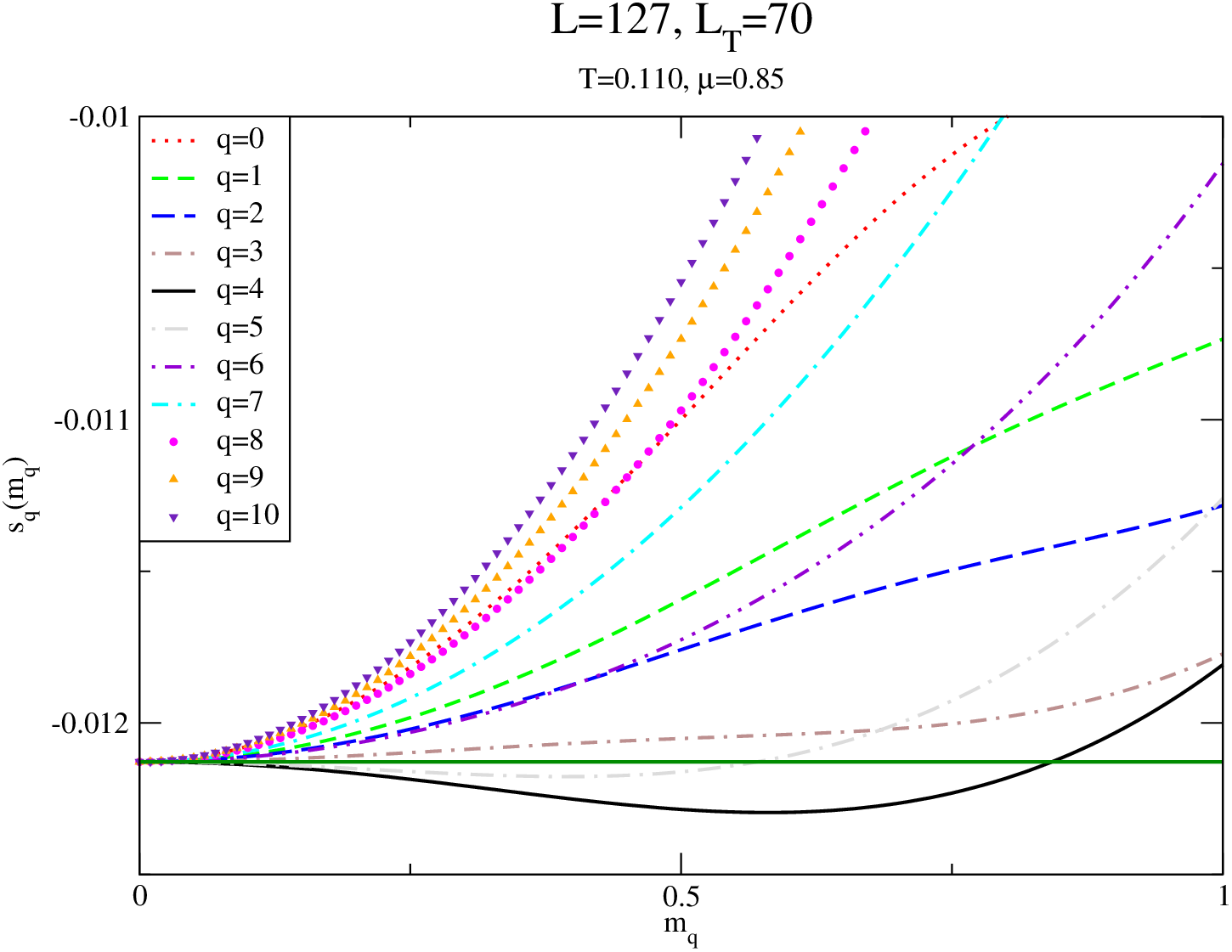}
\includegraphics[scale=0.325]{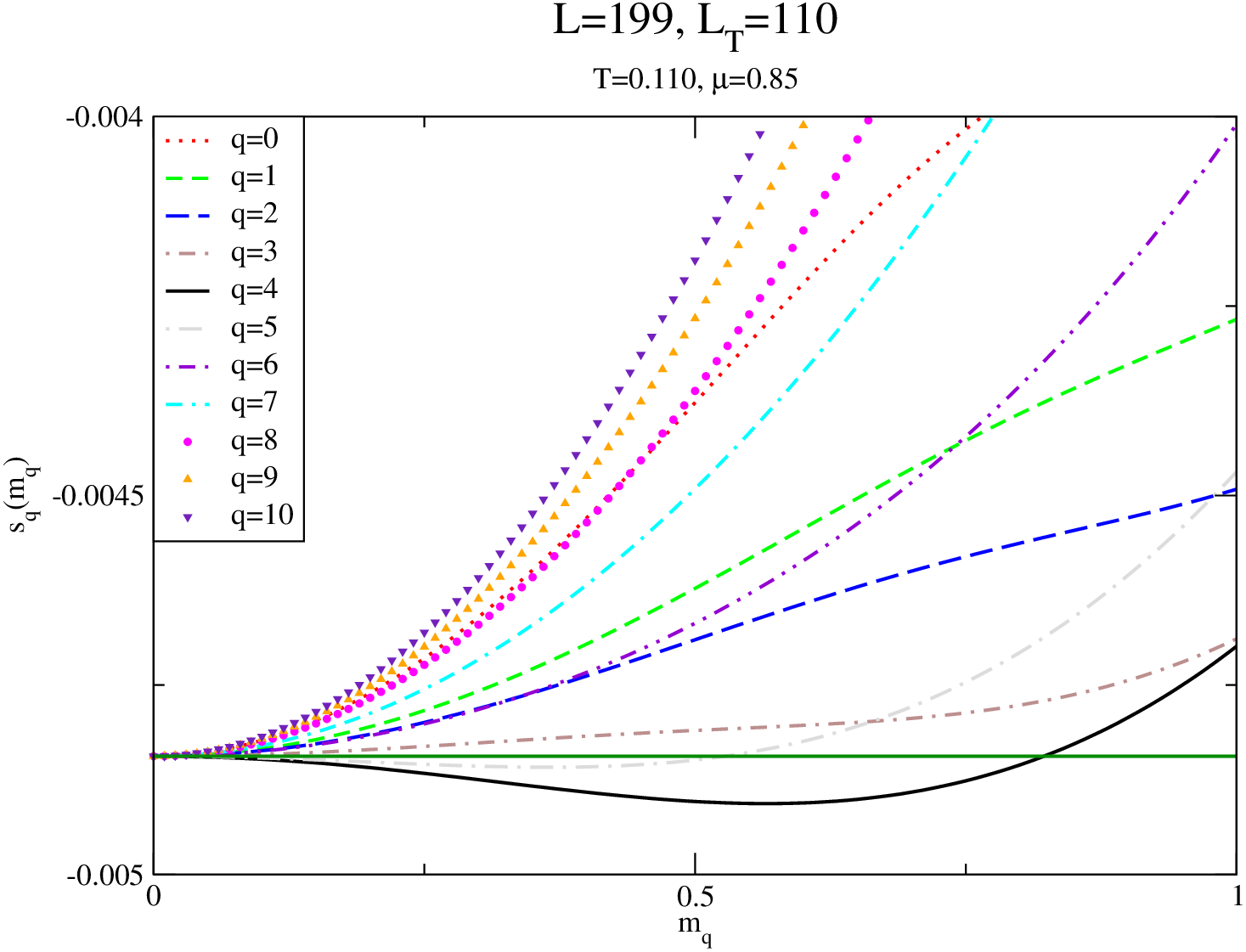}
\caption{ A sample plot of the action density $s_q(m_qm_s;\lambda(m_s),\mu m_s, L, L_T)-s_0(0;0,0,L,L)$
as a function of the condensate, $m_q$, at a chemical potential, $\mu=0.85$ and temperature
$T=0.110$ with an aspect ration of $\tau=\frac{70}{127}$. The two plots correspond to different $(L,L_T)$ with the same aspect ratio. The solid green horizontal line corresponds to the common value of the action density at $m_q=0$.
The y-axis label has suppressed the auxiliary dependences in  \eqn{actden} and \eqn{actdenq}
and the overall subtraction.
}
\eef{actionden2d}

To start off, we plot the action density in \eqn{actden} and \eqn{actdenq} 
as a function of the scalar condensate, $m_q$, at a temperature of $T=0.110$ and
a chemical potential of $\mu=0.85$ for an aspect ratio of $\tau=\frac{70}{127}$  in \fgn{actionden2d}. We have plotted $$s_q(m_qm_s;\lambda(m_s),\mu m_s,L,L_T)
- s_0(0;0,0,L,L)$$ for two different values of $(L,L_T)$ at a same aspect ratio.
The even functions of $m_q$ have the same value at $m_q=0$ shown as a  solid green horizontal line. Both $q=4$ and $q=5$ have
a minimum at the non-zero value of the condensate. 
The minimum of the action over all spatial momenta in the range $q=[0,10]$ occurs at $q=4$ shown using
a solid black line in both plots. Furthermore, there are no discernible lattice spacing effects when $(L,L_T)=(127,70)$ is compared with
$(L,L_T)=(199,110)$.

\bef
\centering
\includegraphics[scale=0.325]{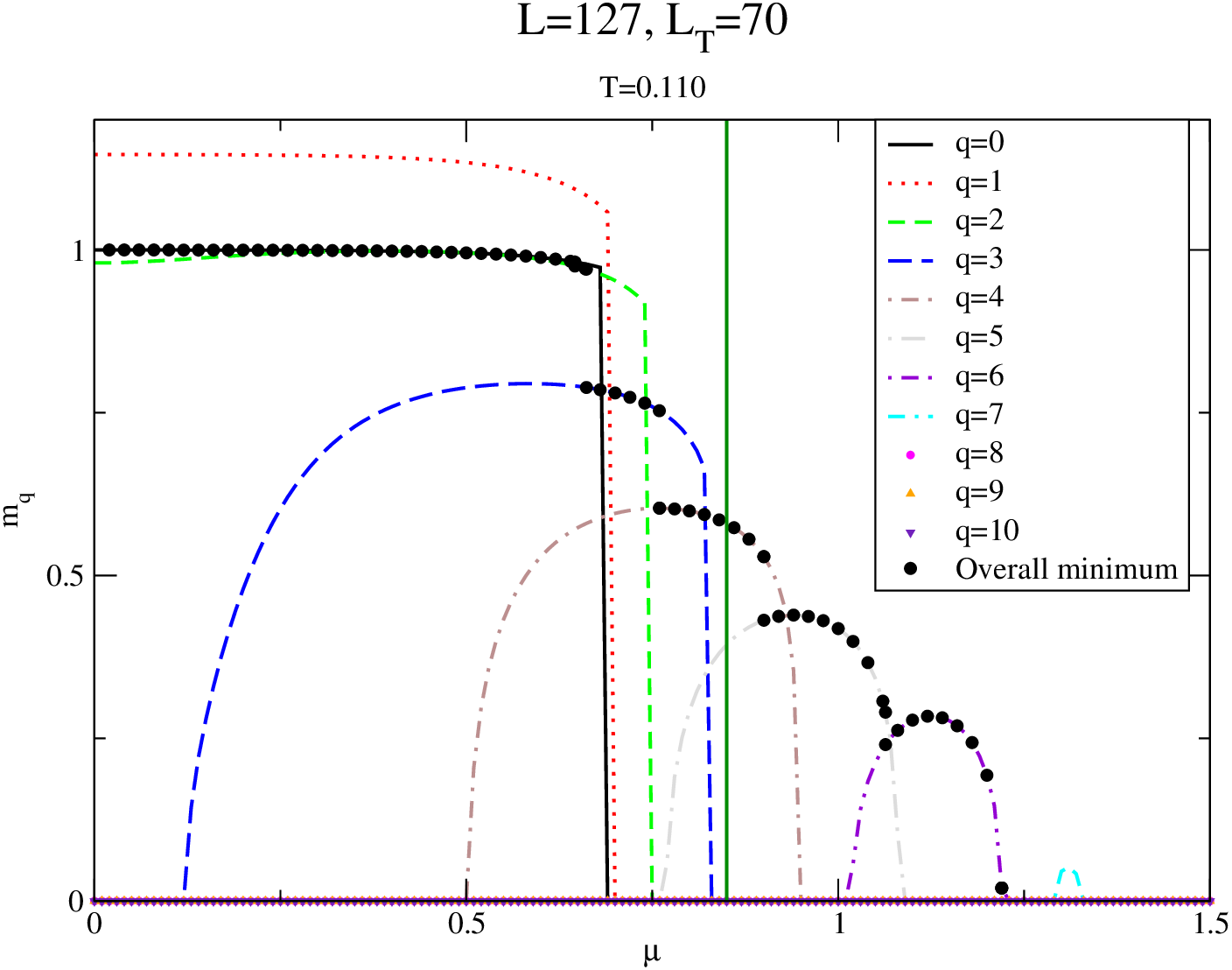}
\includegraphics[scale=0.325]{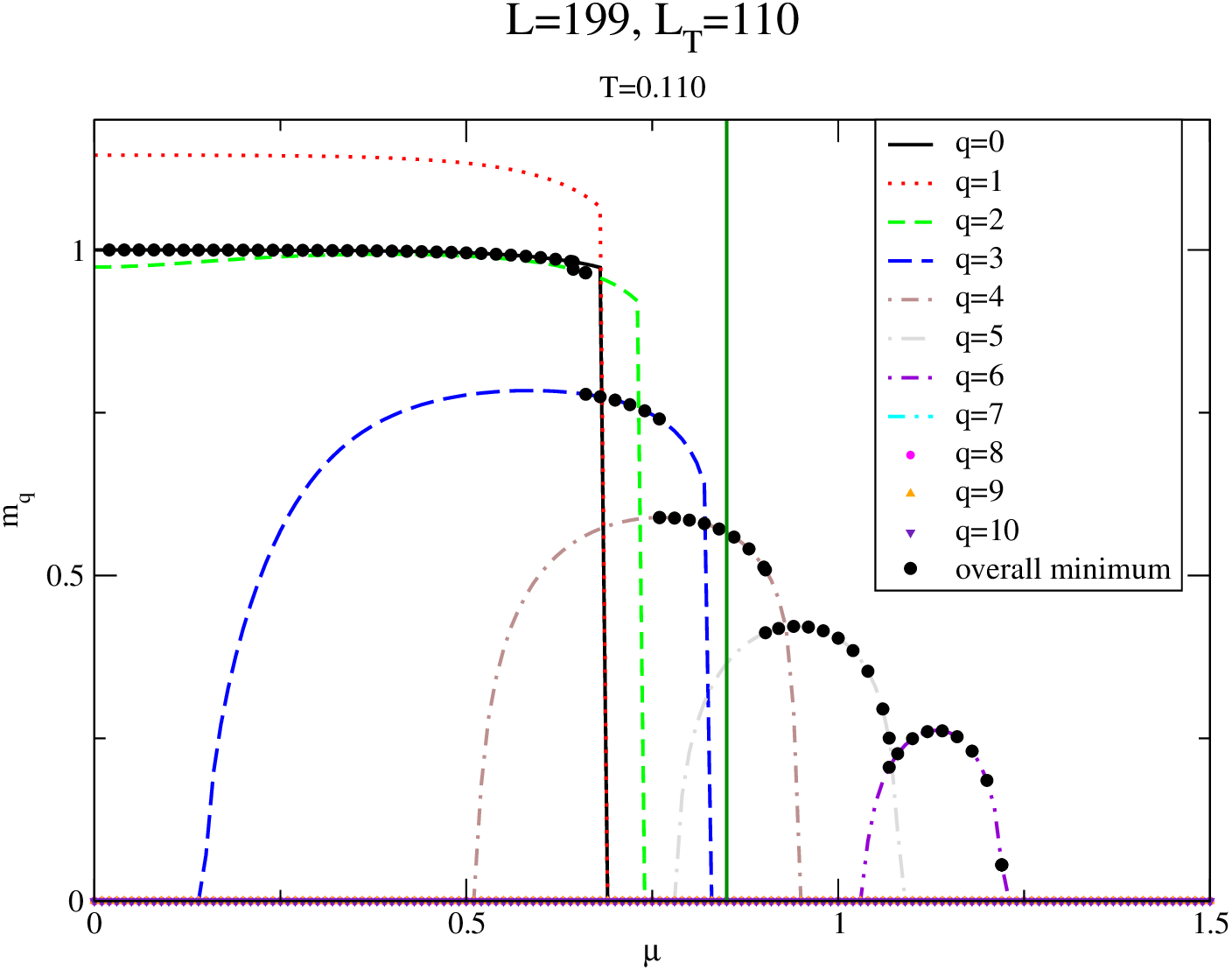}
\caption{ A sample plot of the condensate, $m_q$, as a function of the chemical potential, $\mu$, at
a temperature of $T=0.110$ with an aspect ratio of 
or $\tau=\frac{70}{127}$ . The two plots correspond to different $(L,L_T)$ with the same aspect ratio.
The solid green vertical line corresponds to $\mu=0.85$.
}
\eef{cond2d}

Next, we plot the location of the minimum of the action density for all spatial momenta, $m_q$, in the range $q\in[0,10]$
as a function of the chemical potential, $\mu$, at a temperature of $T=0.110$ for an aspect ratio of $\tau=\frac{70}{127}$ in \fgn{cond2d}. Like in \fgn{actionden2d}, we have shown the results for two different values of
$(L,L_T)$ with the same aspect ratio. To facilitate comparison with \fgn{actionden2d}, we have marked $\mu=0.85$
with a green line where the minimum occurs at $q=4$ when taken over all spatial momenta in the range $q\in[0,10]$.
Note that $q=5$ also gives rise to a non-zero condensate at $\mu=0.85$ but is not favored in comparison to $q=4$.
The points with the black dots in \fgn{cond2d} are the value of the minima taken over all spatial momenta in the range $q\in[0,10]$. 
It is evident from the behavior in \fgn{actionden2d} that this result will not change if the range of spatial momenta is extended. The value of the condensate, $m_0$, in the homogeneous phase at this low a temperature
is essentially independent of $\mu$.
The amplitude of the condensate, $m_q$, decreases with increasing $\mu$; an effect that was observed in \cite{Pannullo:2019prx}. The behavior of the condensate, particularly at $q=5,6$ is not monotonic in the chemical
potential and this is reflective of the contribution to a non-zero condensate at these values of $q$ arising in a finite
window in the chemical potential removed from $\mu=0$.
At this value of temperature only $q\in[0,6]$ contribute to a non-zero value of the condensate. The plots in \fgn{cond2d}  show a transition from a homogeneous condensate phase ($m_0\ne 0$) to
an inhomogeneous condensate phase ($m_2 \ne 0$) at a certain value of the chemical potential. Note that the
minimum does not occur for $q=1$ at any value of the chemical potential.
This is followed
by transitions within the inhomogeneous phase from $m_q \to m_{q+1}$ with $q=2,3,4,5$ at increasing values of
the chemical potential. Finally, there is a transition from the inhomogeneous phase ($m_6\ne 0$) to the symmetric
phase at a certain value of the chemical potential. There are no discernible lattice spacing effects when $(L,L_T)=(127,70)$ is compared with
$(L,L_T)=(199,110)$.
The sequence of transitions within the inhomogeneous phase is the new feature found in this paper as a result of the
analysis at a fixed momentum of the condensate.

\bef
\centering
\includegraphics[scale=0.65]{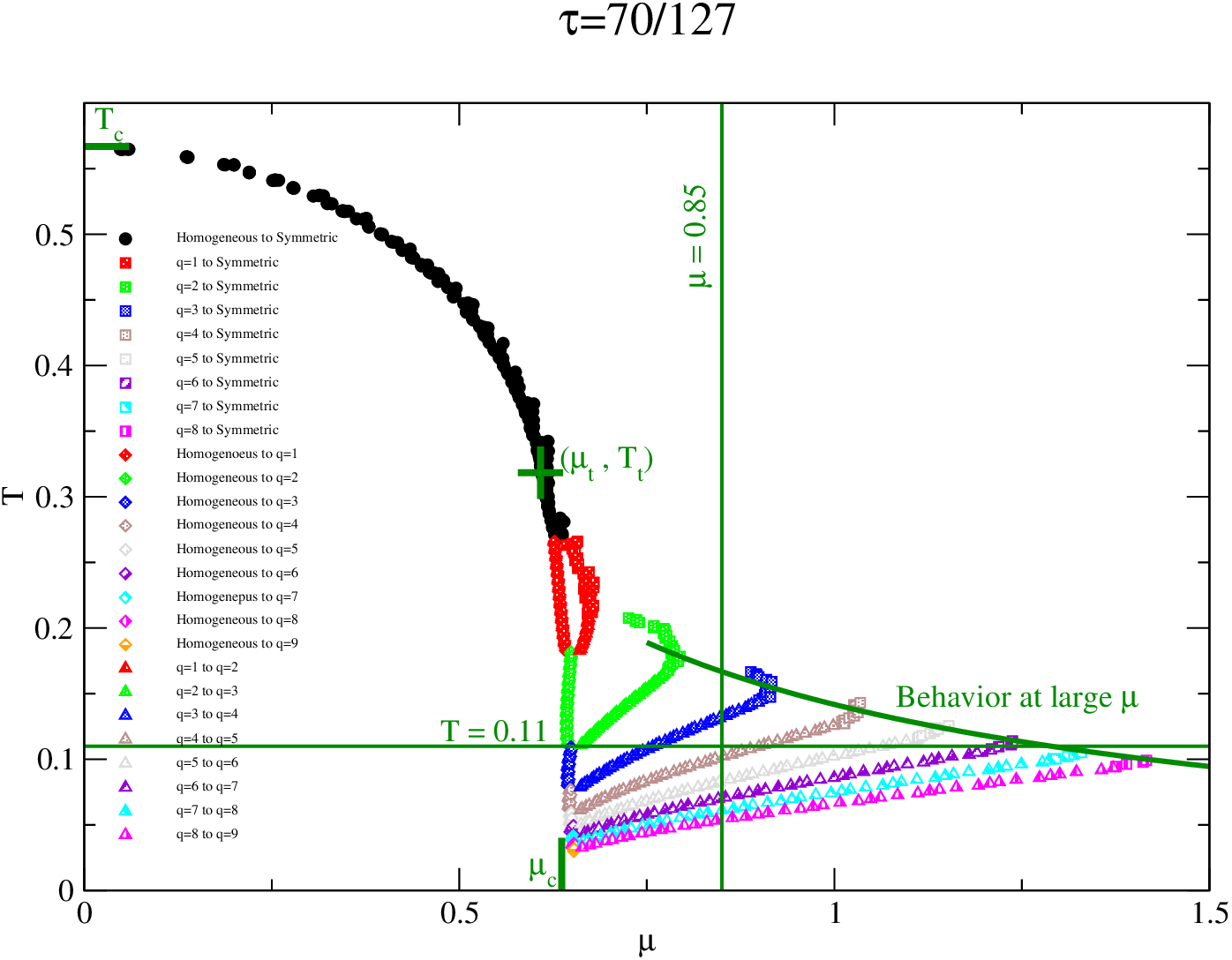}
\caption{The phase diagram at an aspect ratio of $\tau=\frac{70}{127}$.
}
\eef{bigpicture2d}

The full phase diagram will all the features is shown in \fgn{bigpicture2d} for an aspect ratio of $\tau=\frac{70}{127}$.
Four different sets listed in \eqn{lltlist} with the same aspect ratio contributed to
the plot and we see no significant finite lattice spacing effect. We will assume the plot is indicative of the result in the
continuum at this aspect ratio when we discuss its features.
The following types of transitions are present in the figure:
\begin{itemize}
\item Homogeneous phase to symmetric phase.
\item Homogeneous phase to inhomogeneous phase. The momentum in the inhomogeneous phase increases with
decreasing temperature.
\item Inhomogeneous phase to symmetric phase. The momentum in the inhomogeneous phase increases with
decreasing temperature.
\item Phase transitions within the inhomogeneous phase from wavenumber $q$ to wavenumber $q+1$.
\end{itemize}
The transitions from $0$ (homogeneous) to $q$ (inhomogeneous) ; $q$ (inhomogeneous) to $q+1$ (inhomogeneous) and $q$ (homogeneous) to symmetric are smoothly connected. We have marked $\mu=0.85$ and $T=0.110$ in order to make
connection with \fgn{actionden2d} and \fgn{cond2d}. As shown in \fgn{actionden2d}, $(\mu=0.85, T=0.110)$ lies
in the $q=4$ homogeneous phase. As shown in \fgn{cond2d}, the line at $T=0.110$ moves from homogeneous $\to 2 \to 3 \to 4 \to 5 \to 6 \to$ symmetric phase as the chemical potential is increased.

The second order transition at $(0,T_c=\frac{e^C}{\pi})$ 
from the homogeneous phase to the symmetric phase~\cite{Wolff:1985av}
is marked by a solid green tick mark and the transition at this aspect ratio is very close to it. Only in the limit of $\tau\to 0$ do we expect this
transition to approach the solid green tick mark.

One of the discrepancies between the original analysis of the phase diagram in~\cite{Wolff:1985av} and the revised analysis in~\cite{Thies:2003kk}
is the location and nature of the transition at $(0,\mu_c)$. The revised analysis of baryonic matter at zero temperature
supports a second order phase transition  at $\mu_c=\frac{2}{\pi}$ where as the original analysis assuming a homogeneous condensate supports
a first order phase transition at $\mu_c=\frac{1}{\sqrt{2}}$. We have marked $\mu_c=\frac{2}{\pi}$ by a solid green tick mark and our analysis assuming a condensate with a fixed momentum suggests a transition from the homogeneous phase to an inhomogeneous phase. Since 
we are studying the model at a fixed aspect ratio, the model is in an infinite spatial extent at zero temperature. Since we are studying it with a lattice regularization, our finite lattice spacing effects get larger as we approach zero temperature. In spite of this, our estimate of the transition point is close to $\mu_c=\frac{2}{\pi}$ and the wavenumber in the inhomogeneous phase grows as we decrease $T$.

The analysis in~\cite{Thies:2003kk} shows that the model remains in the inhomogeneous phase at zero temperature for all $\mu > \mu_c$.
The prediction for the phase boundary from the inhomogeneous phase to the symmetric phase is 
\be
T_{\rm crit} = \frac{e^C}{4\pi\mu}
\ee
and this is shown as a solid green curve in \fgn{bigpicture2d} for $\mu \ge \frac{3}{4}$. In our analysis, the transition from the inhomogeneous phase to symmetric phase is shown by square dots with various patterns with the different patterns corresponding to different values of wavenumber, $q$, from which
the system goes into the symmetric phase. The momentum at the transition increases with $\mu$ and the location of the transition is
quite consistent with the asymptotic behavior. 

The tricritical point at $(\mu_t=0.60822,T_t=0.31833)$ in the original analysis~\cite{Wolff:1985av} is interpreted as a Lifshitz point in~\cite{Thies:2003kk}. This point is marked with a solid green cross-wire in \fgn{bigpicture2d}. This point lies in the second order line in our
analysis that separates the homogeneous phase from the symmetric phase. The region in the inhomogeneous phase with wavenumber, $q=1$,
is bounded by red diamonds (homogeneous to $q=1$), red triangles ($q=1$ to $q=2$) and red squares ($q=1$ to symmetric)
in our analysis. The momentum associated with the intersection of the red diamonds with red squares 
 has a momentum of $\frac{2\pi}{\ell}$ and this occurs at a certain value of $\mu$ and $T$ that is close to the tricritical point.
If the tricritical point is indeed a Lifshitz point, we expect our estimate to be different since our aspect ratio is not zero.
In order to understand this point and other effects of a non-zero aspect ratio, we have shown the phase diagram at the  six other
values of aspect ratios listed in \eqn{taulist} in \fgn{phase-diagram-2d}. 
Four different values of $(L,L_T)$ listed in \eqn{lltlist} contributed to each one of the aspect ratios and like in the case of $\tau=\frac{70}{127}$ shown in \fgn{bigpicture2d}, we do not see any significant finite lattice
spacing effects.

\bef
\centering
\includegraphics[scale=0.325]{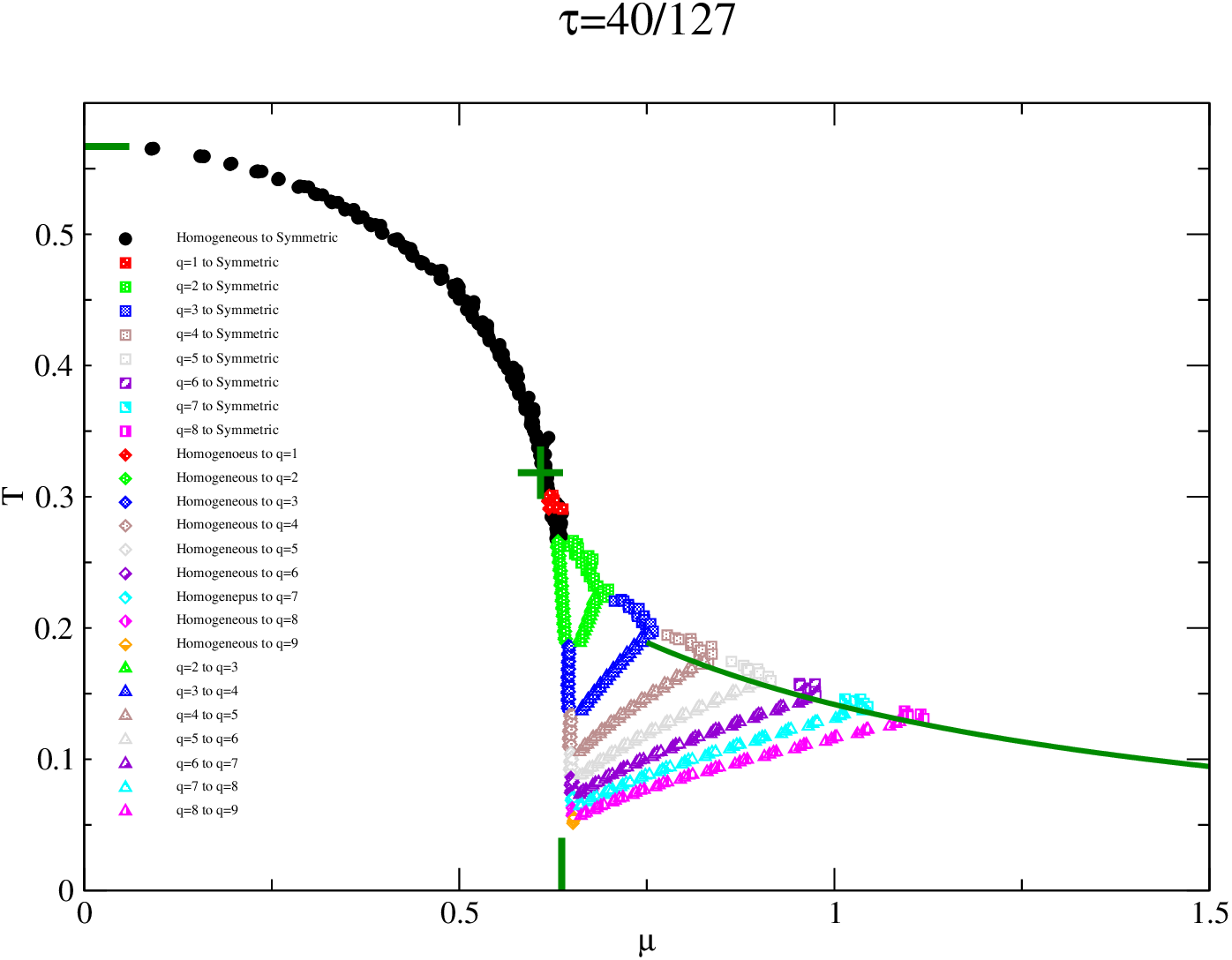}
\includegraphics[scale=0.325]{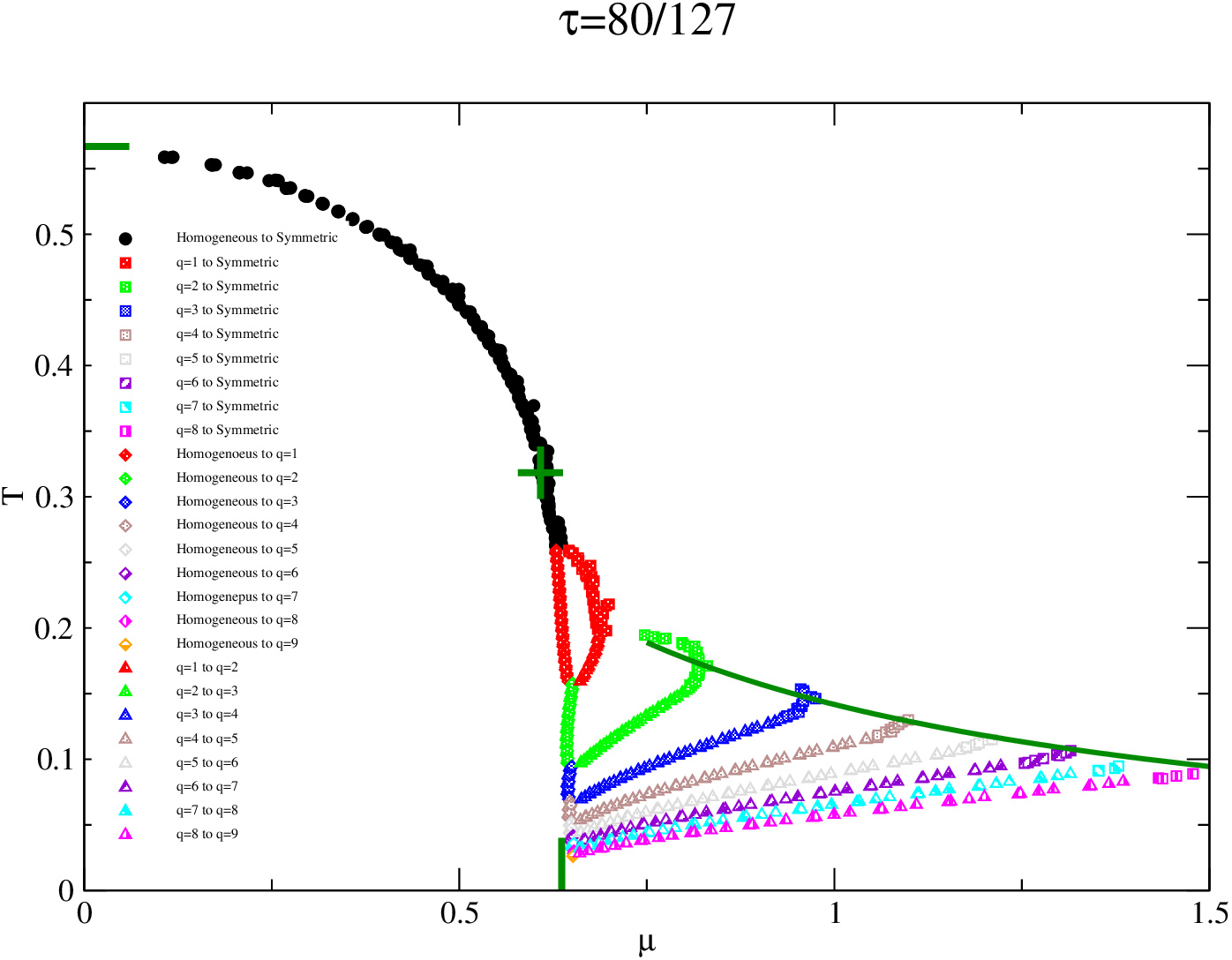}
\includegraphics[scale=0.325]{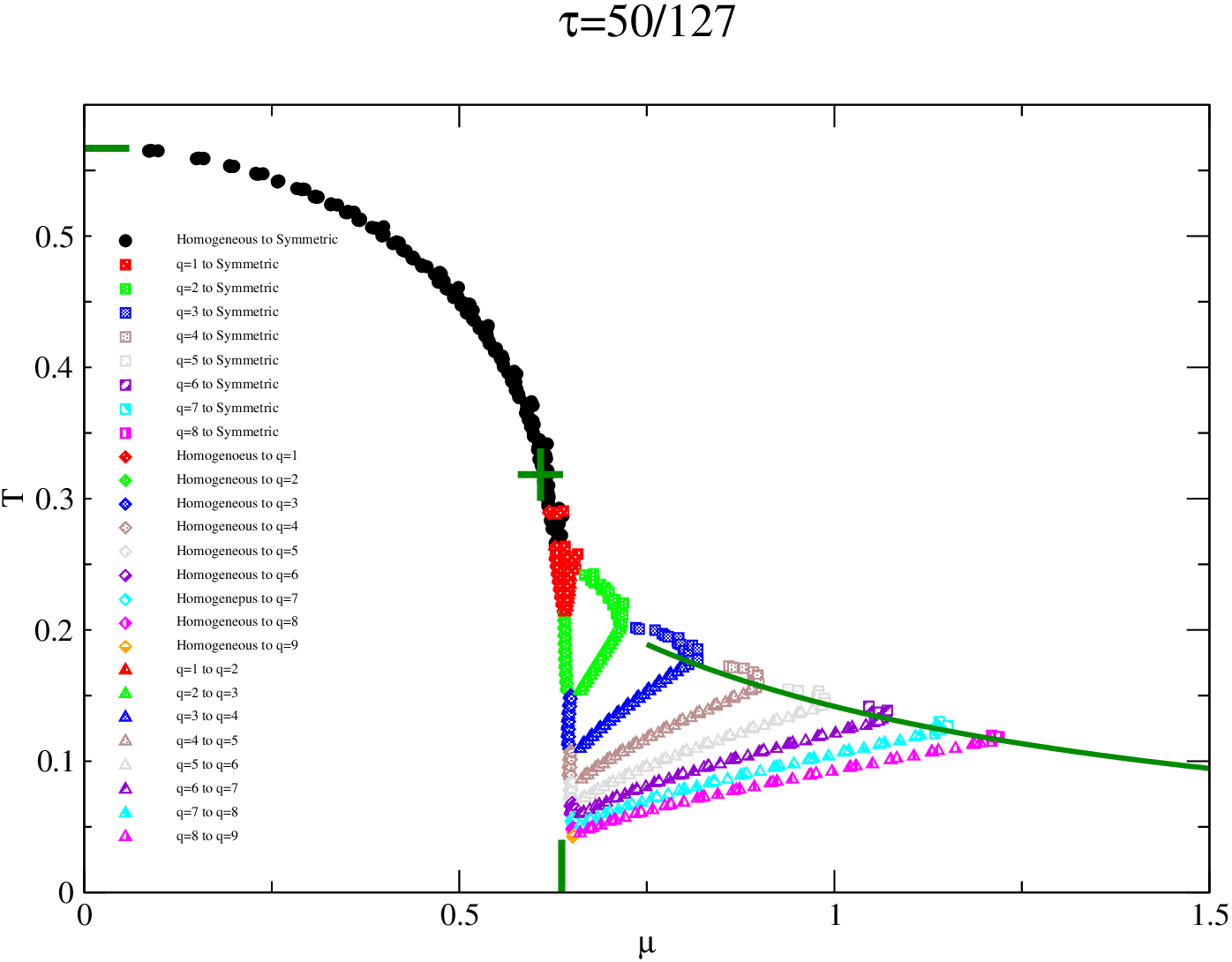}
\includegraphics[scale=0.325]{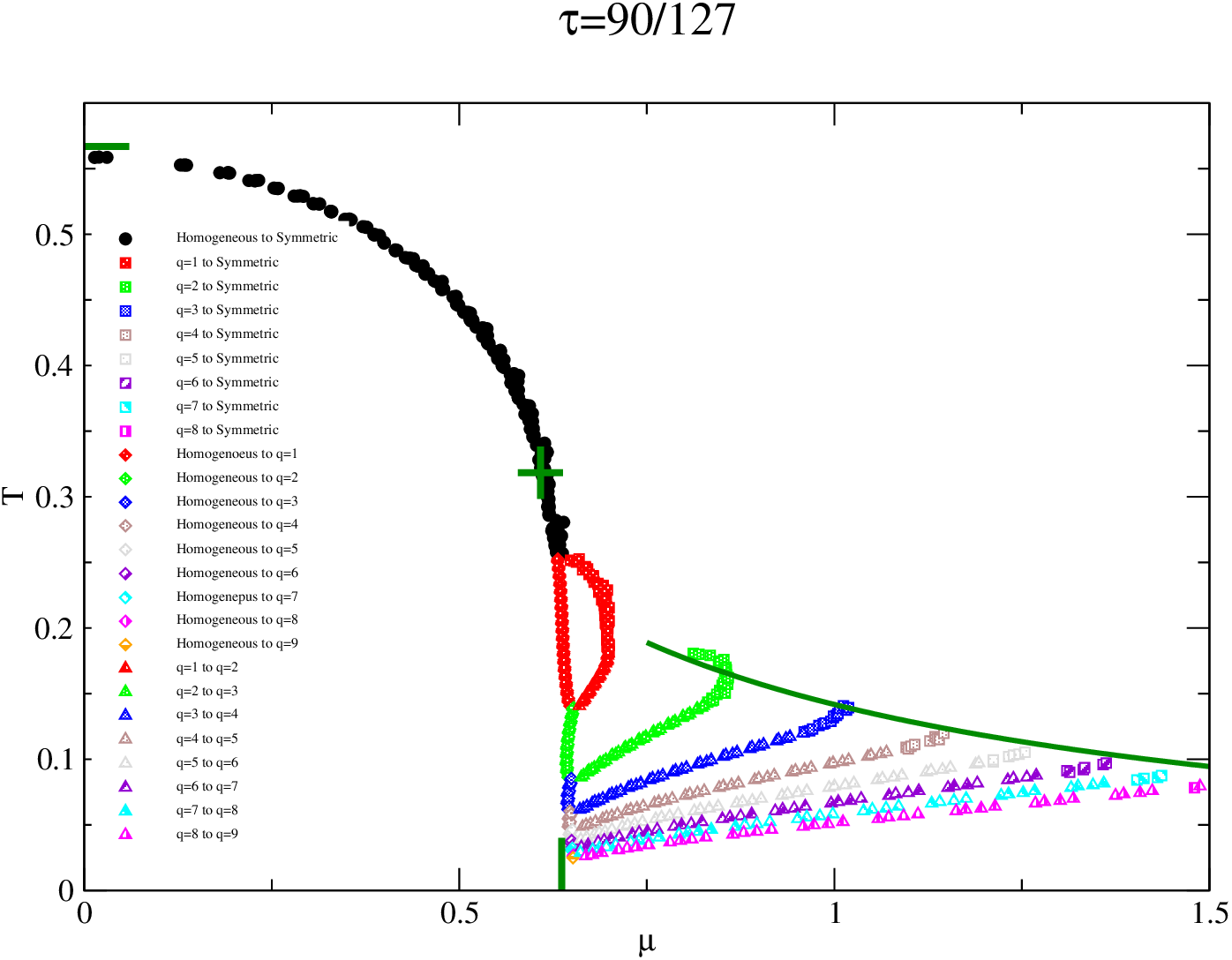}
\includegraphics[scale=0.325]{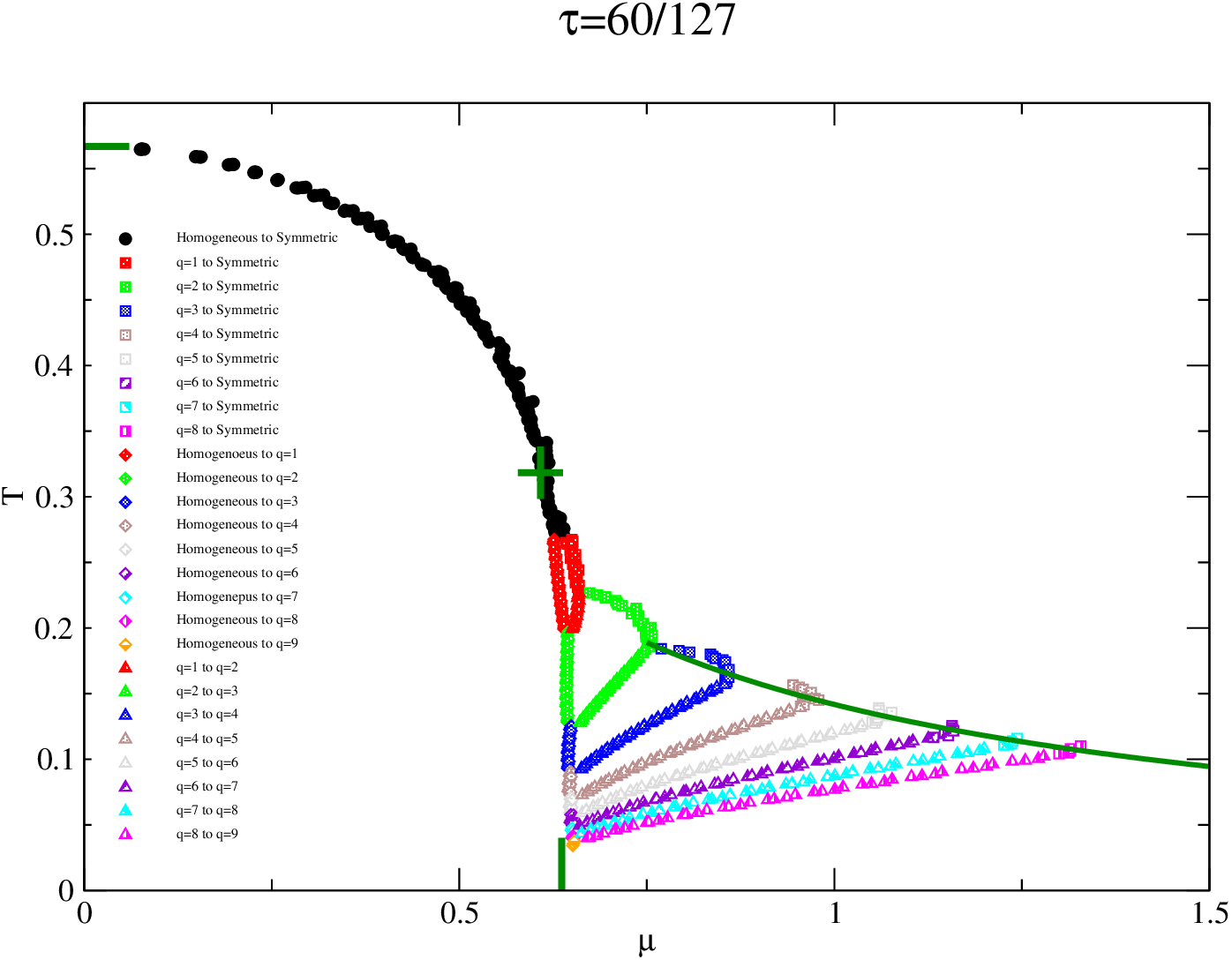}
\includegraphics[scale=0.325]{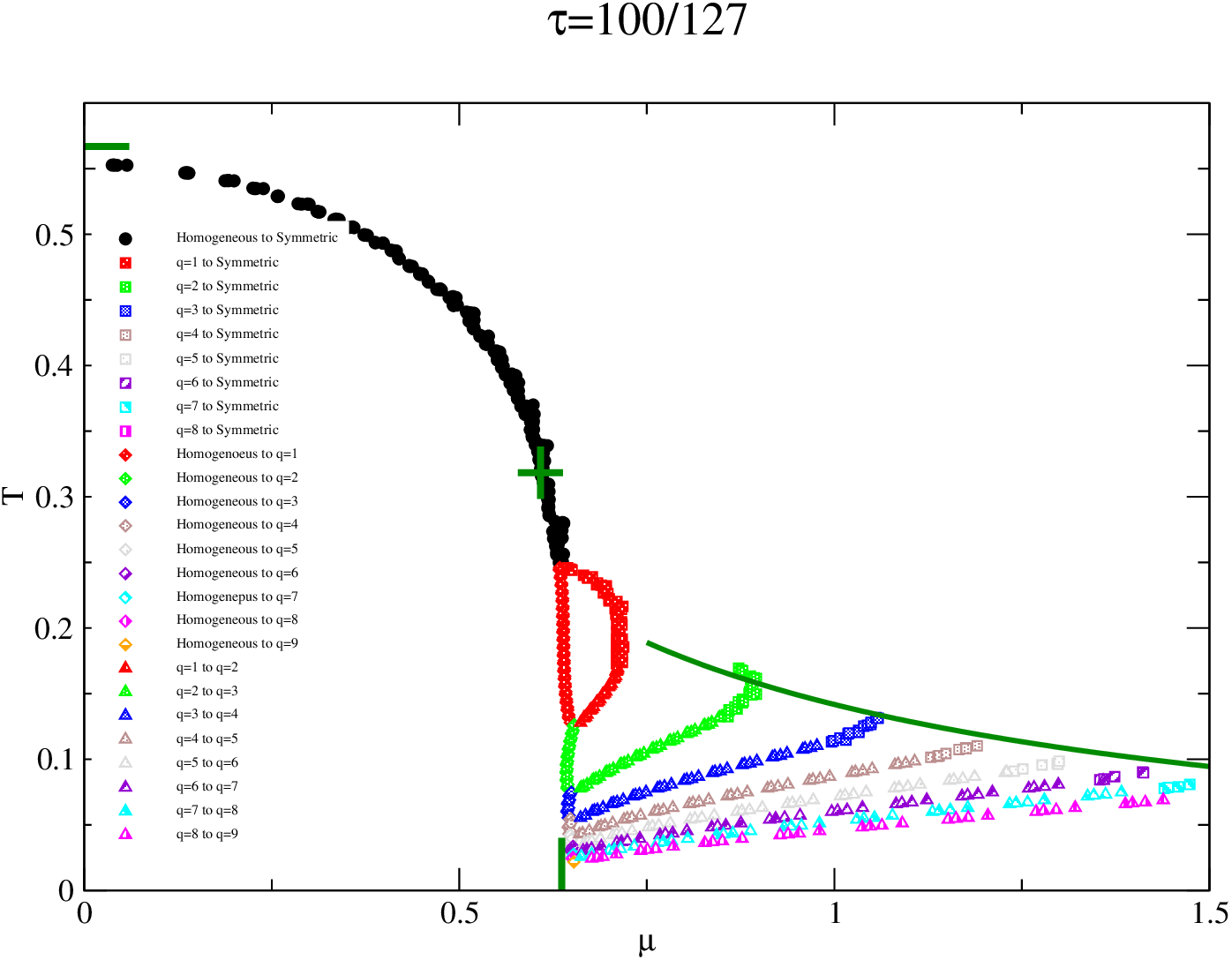}
\caption{ The phase diagram for the two dimensional Gross-Neveu model in the large $N$ limit on a finite torus with six different aspect ratios.
}
\eef{phase-diagram-2d}

With regard to the tricritical point, we see two effects with increasing aspect ratio. The region with wavenumber, $q=1$,
grows with increasing aspect ratio and the region moves away from $(\mu_t,T_t)$. The aspect ratio has changed by a factor of $\frac{5}{2}$
over the entire range but the temperature associated with the intersection of the red diamonds with the red squares has not changed
proportionately. This implies that the spatial extent, $\ell$, associated with the intersection point has increased with decreasing aspect
ratio suggesting that the momentum becomes smaller with decreasing $\tau$. This is consistent with the intersection point
approaching a Lifshitz point in the limit of zero aspect ratio. With regard to the critical temperature, $T_c$, at zero chemical potential,
we see that the value does not shift much with the aspect ratio and we see a significant deviation 
toward a lower critical temperature only for $\tau > \frac{70}{127}$. With regard to the critical chemical potential, $\mu_c$, at zero temperature,
we do not see any significant change since the spatial extent is approaching infinity at all aspect ratios when we approach zero temperature.
To quantify these statements, we plot our estimate of $(\mu_t,T_t)$, $\mu_c$ and $T_c$ as a function of the aspect ratio in \fgn{scaling-2d}.
The lack of smooth behavior as a function of $\tau$ is due to our estimates arising from using discrete steps in $T$ and $\mu$ to produce
our points in the phase diagram. Since we only wish to show consistency with previous results, we perform a simple linear regression of our
results and compare them with the previous results~\cite{Thies:2003kk} marked by circular dots. Our estimate for $\mu_c$ might be considered
to be a bit higher but the rest of estimates are essentially consistent with the previous results.

\bef
\centering
\includegraphics[scale=0.65]{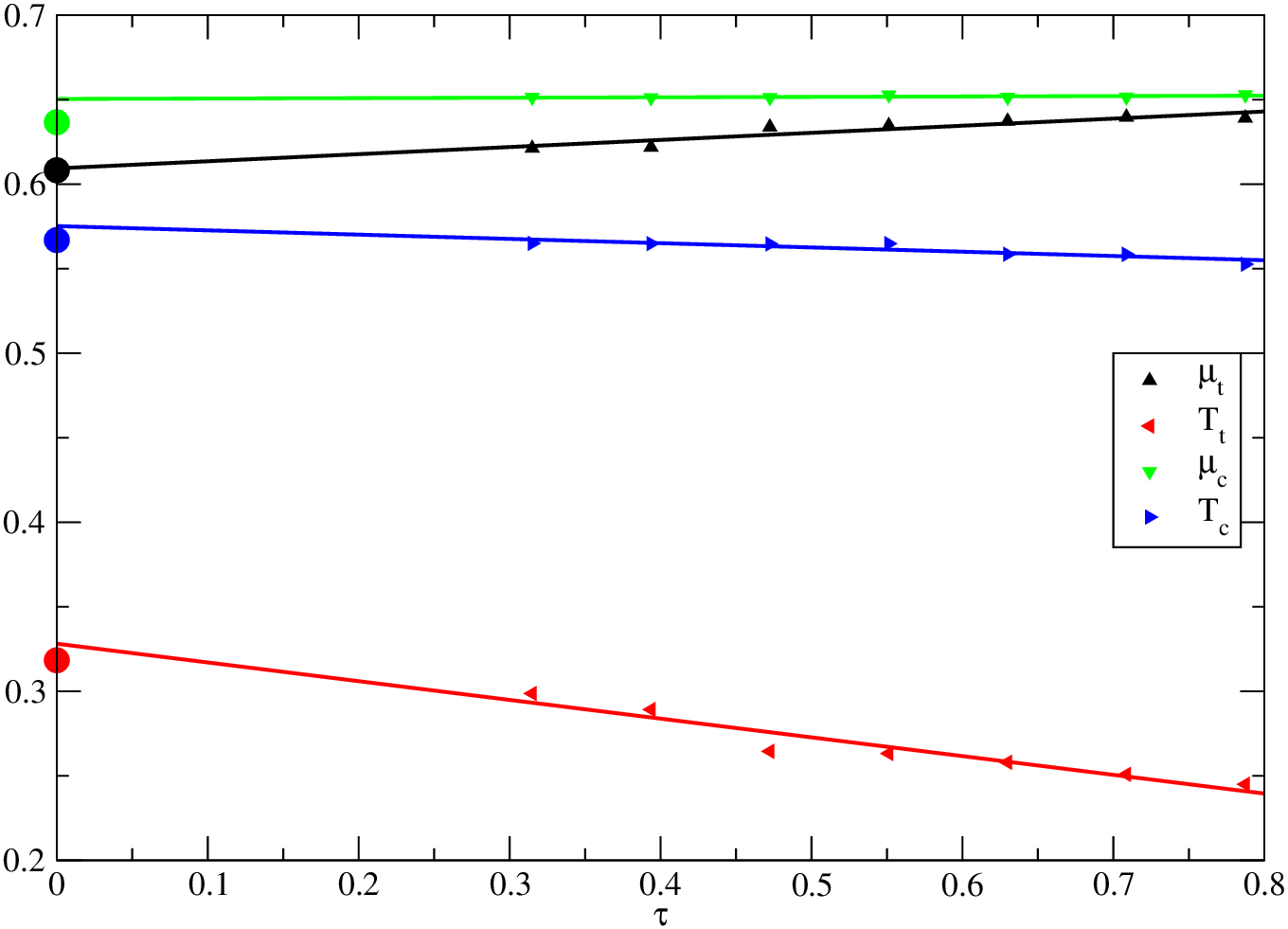}
\caption{ Approach of our estimate of $(\mu_t,T_t)$, $\mu_c$ and $T_c$ to aspect ratio of zero is compared with the analysis in~\cite{Thies:2003kk}.
}
\eef{scaling-2d}

The transition from the inhomogeneous phase to the symmetric phase has two main features. There is a wavenumber associated with
each value of $\mu$ and it grows with increasing $\mu$. Smaller wavenumbers show a significant curvature in the plot of the phase diagram
and this is a effect due to the finite aspect ratio. We define the momentum associated with the wavenumber as
\be
p(\mu) = \frac{\pi q(\mu)}{\ell} = \frac{\pi q(\mu)}{Lm_s}.
\ee
This definition is off by a factor of two from \eqn{sigmom} but is consistent with~\cite{Thies:2003kk}. The plot of $p(\mu)$ is shown in \fgn{mom-vs-muT} for all seven aspect ratios in the left panel. The value of
the chemical potential, $\mu_t$, associated with the tricritical point is shown as a vertical line and the expected asymptotic behavior
of $p(\mu)=\mu$ is also shown as a black solid line. The curvature seen in the phase diagram at small wavenumbers is also seen in this plot.
The asymptotic behavior is linear and is quite possibly consistent with $p(\mu)=\mu$ -- we should note that the results from the larger $\mu$
values come from larger values of the aspect ratio, $\tau$, and we only expect consistency with the asymptotic behavior as $\tau\to 0$.
We do not see any point below $\mu_t$ and this is just a consistency check of our estimate of the tricritical point in \fgn{scaling-2d}. Like
in~\cite{Thies:2003kk} we see the momentum drop as we come close to $\mu_t$ but we see effects of finite aspect ratio.

There is also a wavenumber in the phase diagram that continuously increases as $T$ increases along the critical line that separated the
homogeneous phase from the inhomogeneous phase. Defining the momentum associated with the wavenumber as
\be
p(T) = \frac{2\pi q(T)}{\ell} = \frac{2\pi q(T)}{Lm_s},
\ee
 we plot $p(T)$ for all seven aspect ratios in the right panel of \fgn{mom-vs-muT}. The value of the temperature, $T_t$,
associated with the tricritical point is shown as a vertical line. The plot for each aspect ratio is a set of lines that reduce in the length
as $T$ decreases and seems to converge to $p=1$ in the zero temperature limit. If this is indeed the case in the limit of the aspect ratio
going to zero, our analysis at fixed momentum suggests this transition is from zero momentum to a non-zero momentum implying two different order parameters
on either side of the transition.

\bef
\centering
\includegraphics[scale=0.325]{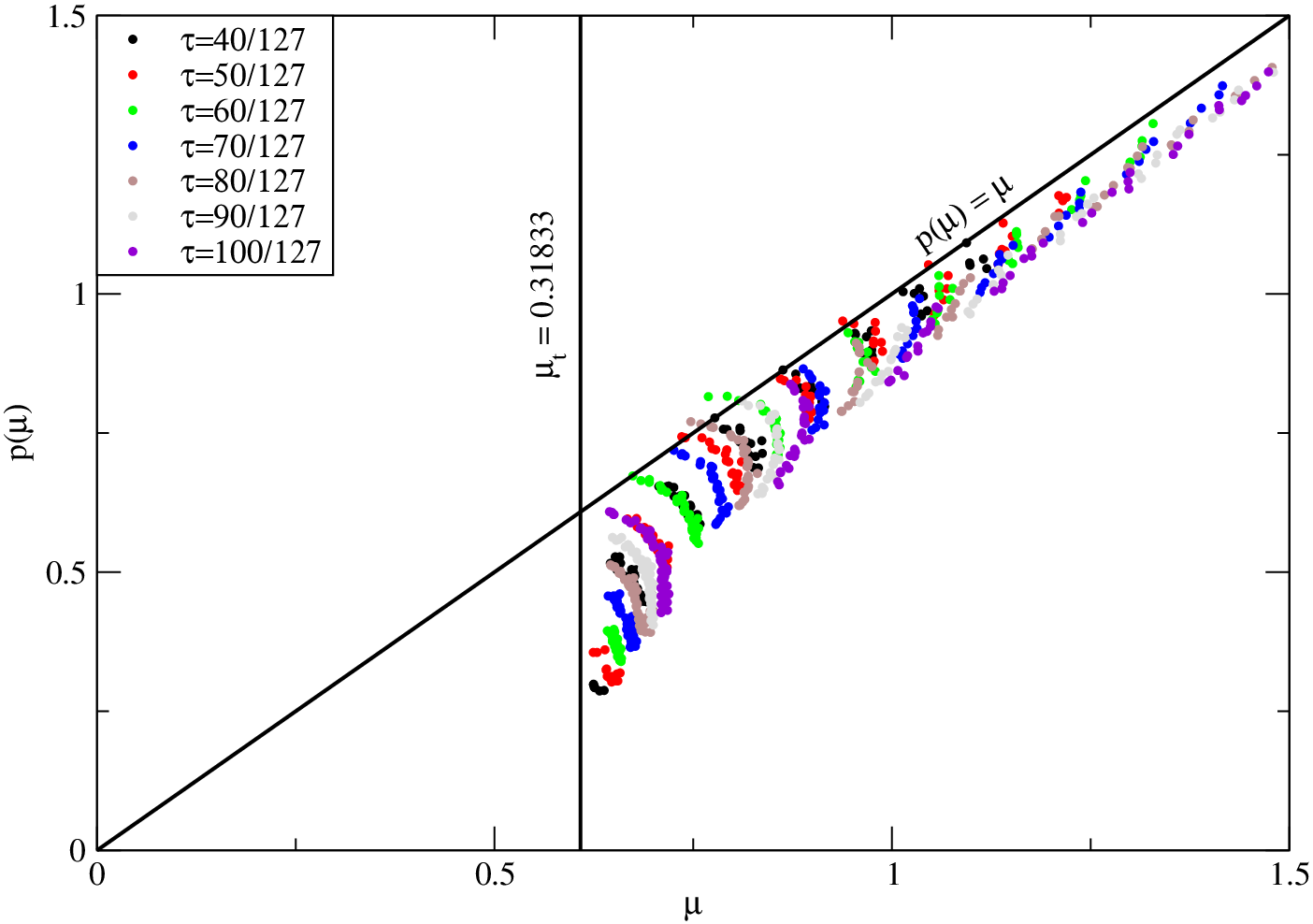}
\includegraphics[scale=0.325]{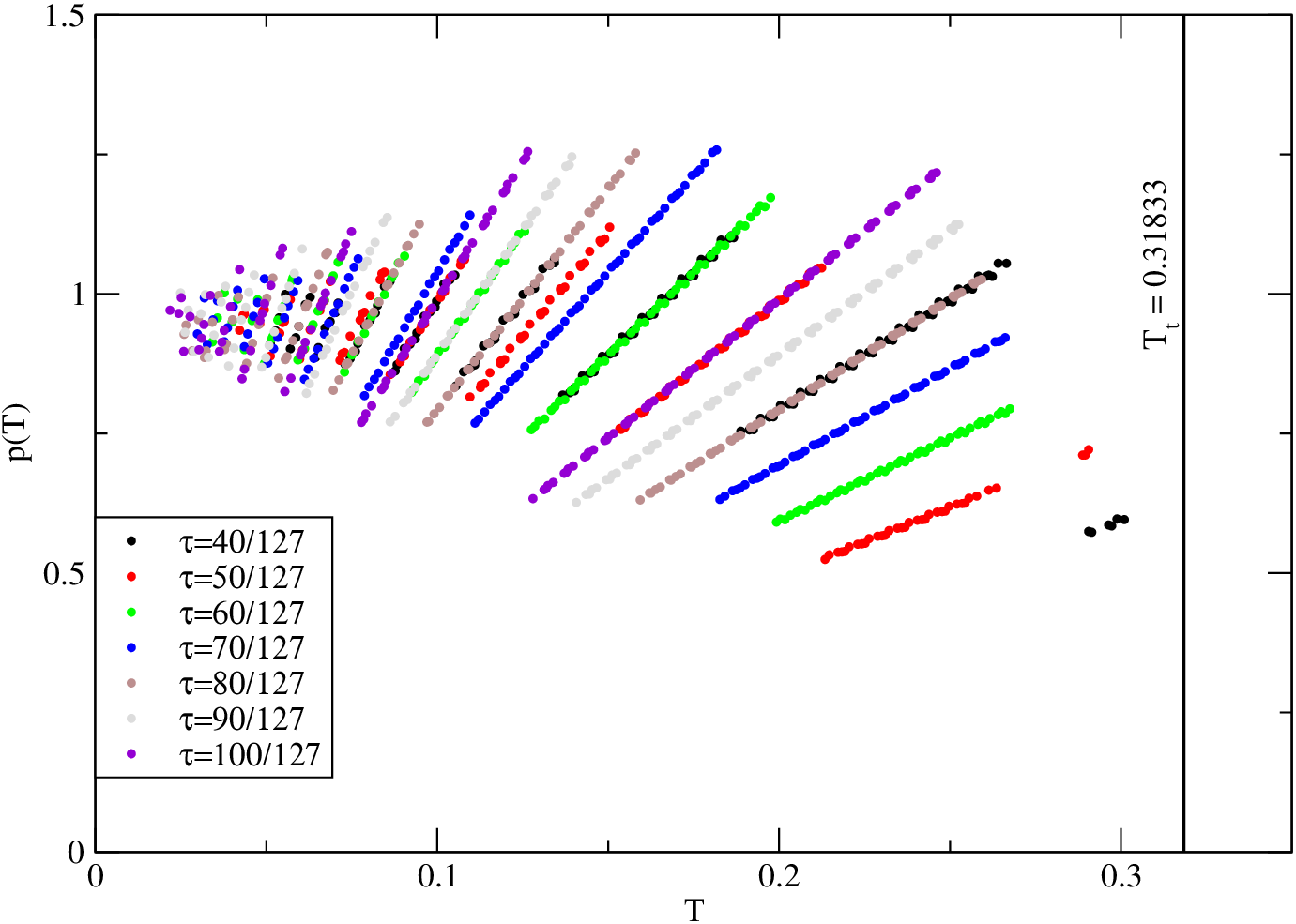}
\caption{ The left panel shows the momentum as a function of the chemical potential along the critical line that separates the inhomogeneous phase from the symmetric phase. The right planel shows the momentum as a function of the temperature along the critical line that separated the homogeneous phase from the inhomogeneous phase.
}
\eef{mom-vs-muT}

\subsection{Three dimensional model}

The analysis with a homogeneous condensate in~\cite{Klimenko:1987gi,Rosenstein:1988dj} shows a line of second order transition given by
\be
\frac{1}{T} = \ln \left [ 2 + 2 \cosh \frac{\mu}{T} \right].\label{critical-3d}
\ee
Unlike a similar analysis in two dimensions~\cite{Wolff:1985av}, there is no tri-critical point in three dimensions. This could be
the reason why an inhomogeneous condensate of a specific form in~\cite{Urlichs:2007zz} was not favored instead of a homogeneous condensate.
A stability analysis on the lattice~\cite{Winstel:2019zfn} did find a region of instability for small temperature and large density.
Since we are also performing an analysis on the lattice, we will find evidence in support of a phase with an inhomogeneous condensate.
But, we will show that unlike the situation in the two dimensional model, the region with an inhomogeneous condensate will
shrink as the lattice spacing is decreased. This suggests that only a phase with a homogeneous condensate survives the continuum limit.
Our analysis will parallel the one performed for the two dimensional model. We will only show results for three aspect ratios, namely,
$\tau=\frac{40}{127},\frac{70}{127},\frac{100}{127}$. To begin, we assume that there is only a phase with a homogeneous condensate
and  a symmetric phase and compare the phase diagram at a fixed aspect ratio with \eqn{critical-3d} in \fgn{bigpicture3d}. Each aspect
ratio has results from four different choices of $(L,L_T)$ listed in \eqn{lltlist}. We see finite lattice spacing effects and non-zero aspect ratio
effects in the plot. Since the values of $L_T$ are smaller for smaller aspect ratios, we see larger lattice spacing effects at $\tau=\frac{40}{127}$.
At temperatures close to $T_c=2\ln 2$ and below, the non-zero aspect ratio effects are small for $\tau=\frac{70}{127}$ and $\tau=\frac{40}{127}$
but $\tau=\frac{100}{127}$ shows significant effect due to non-zero aspect ratio. This behavior is similar to the one in the two dimensional
model.
For the rest of the discussion, we will focus on the
region inside the solid green box bounded by $T \in [0.09,0.21]$ and $\mu \in [1,1.06]$ where we see effects due to finite lattice spacing and
non-zero aspect ratio.

\bef
\centering
\includegraphics[scale=0.65]{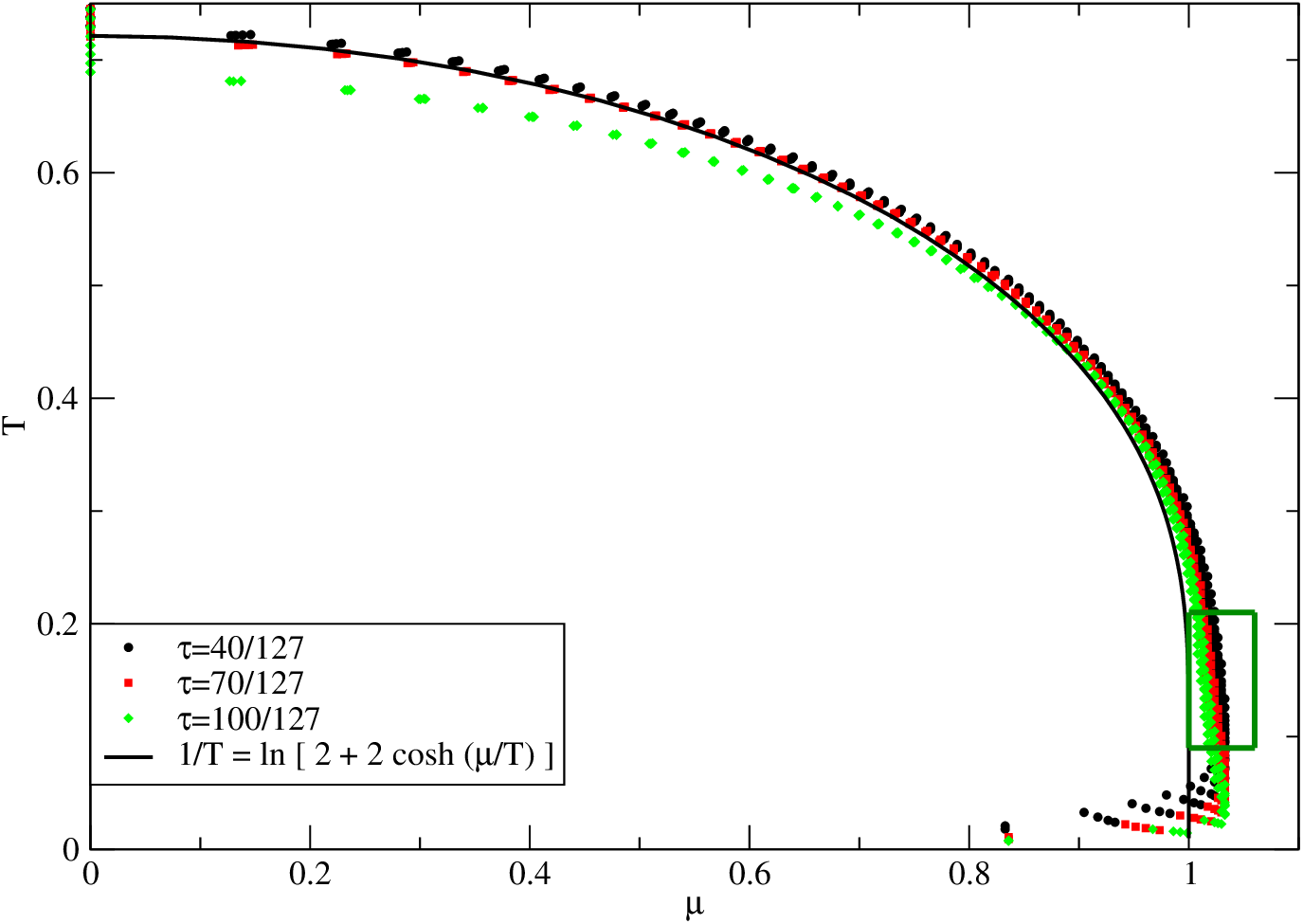}
\caption{ The phase diagram for three different aspect ratios are compared to \eqn{critical-3d} assuming
that only a homogeneous and a symmetric phase exist in the continuum.
}
\eef{bigpicture3d}

\bef
\centering
\includegraphics[scale=0.325]{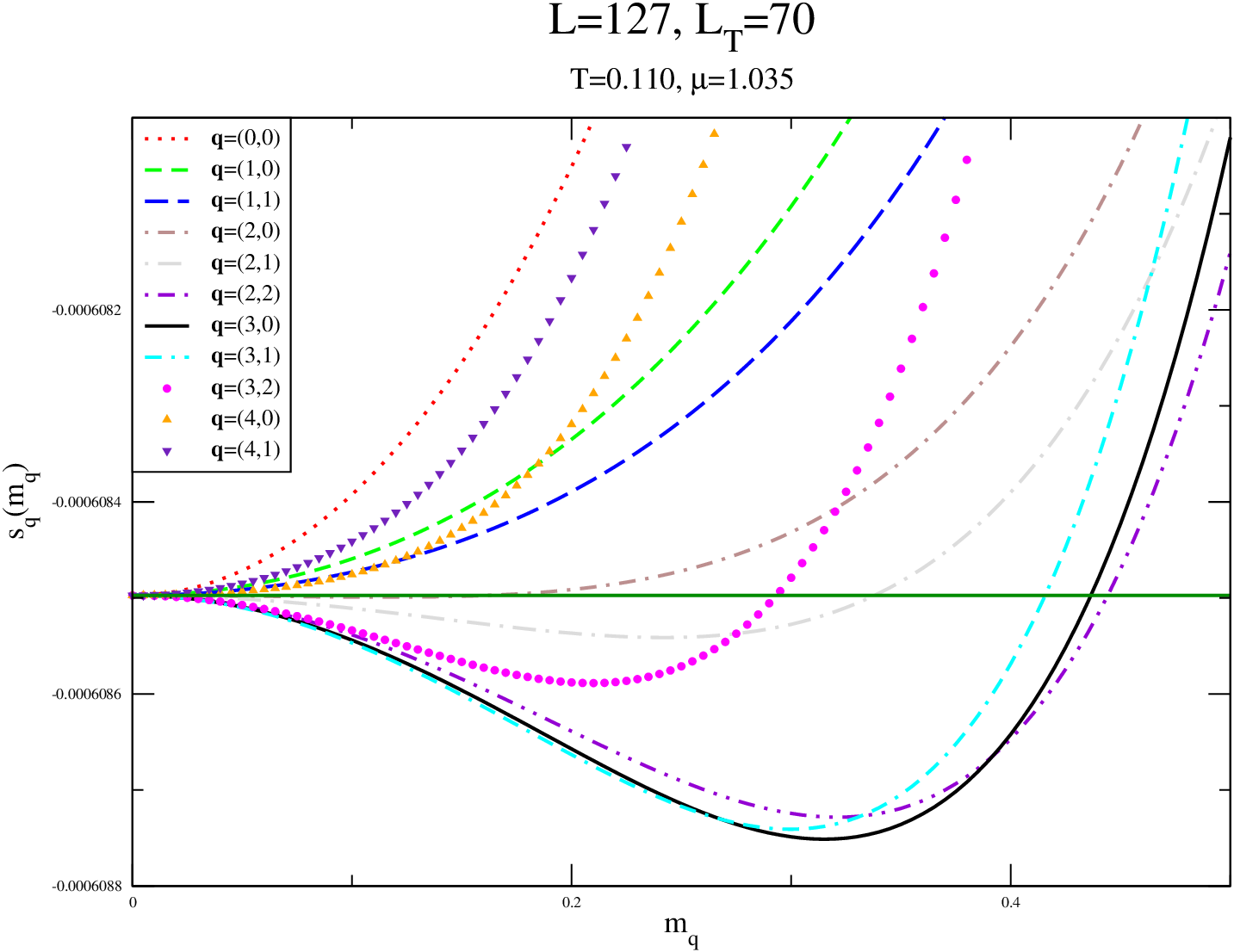}
\includegraphics[scale=0.325]{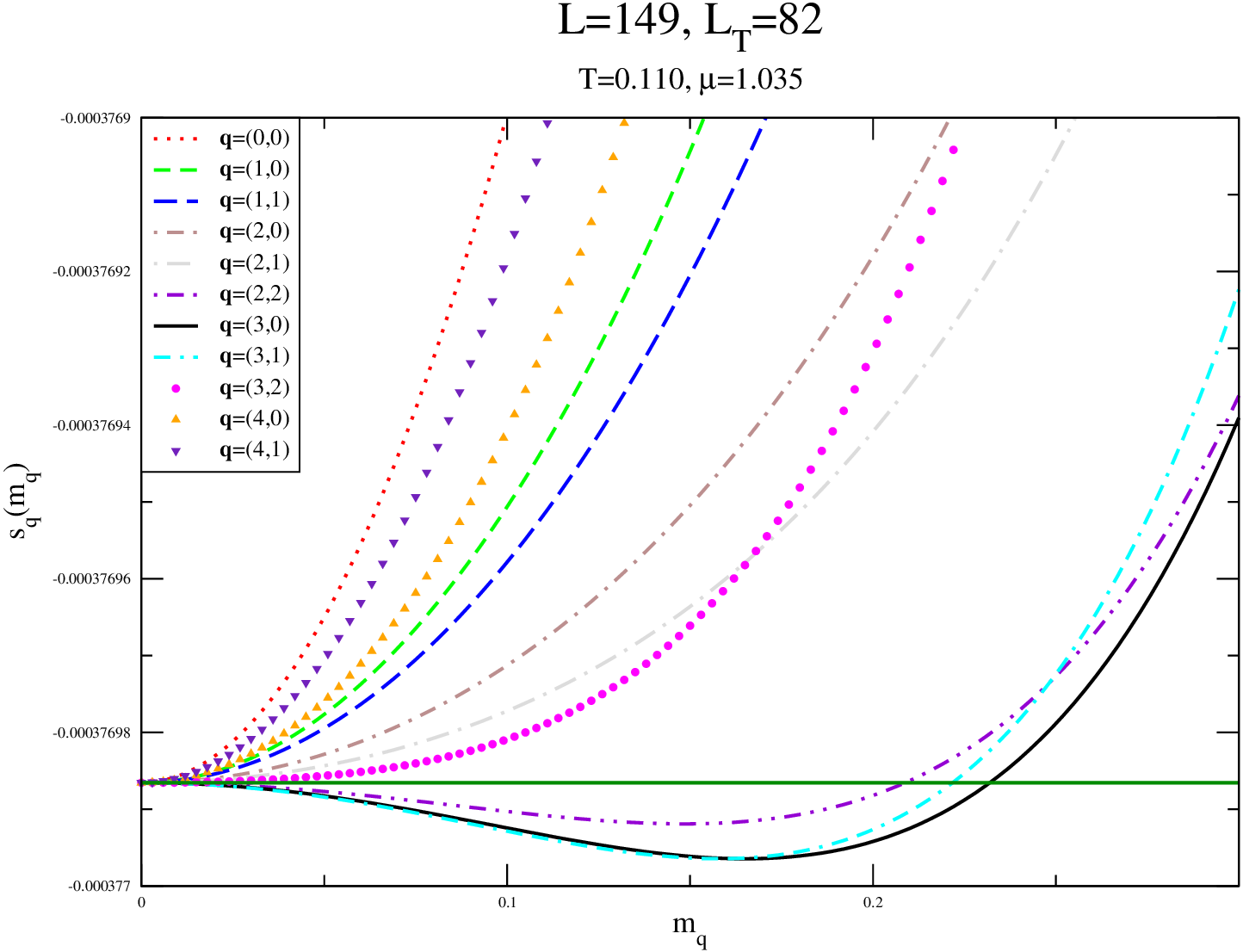}
\includegraphics[scale=0.325]{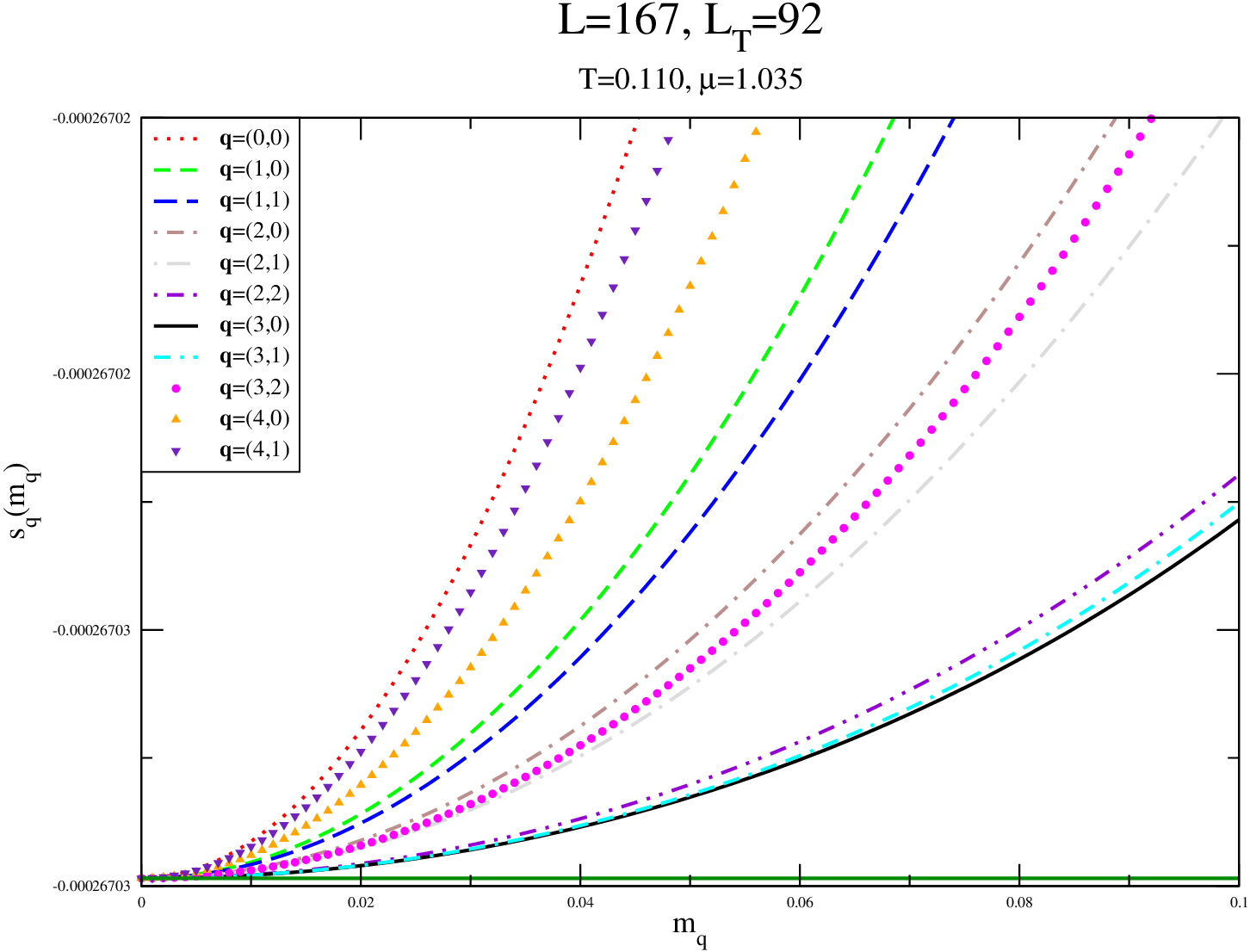}
\includegraphics[scale=0.325]{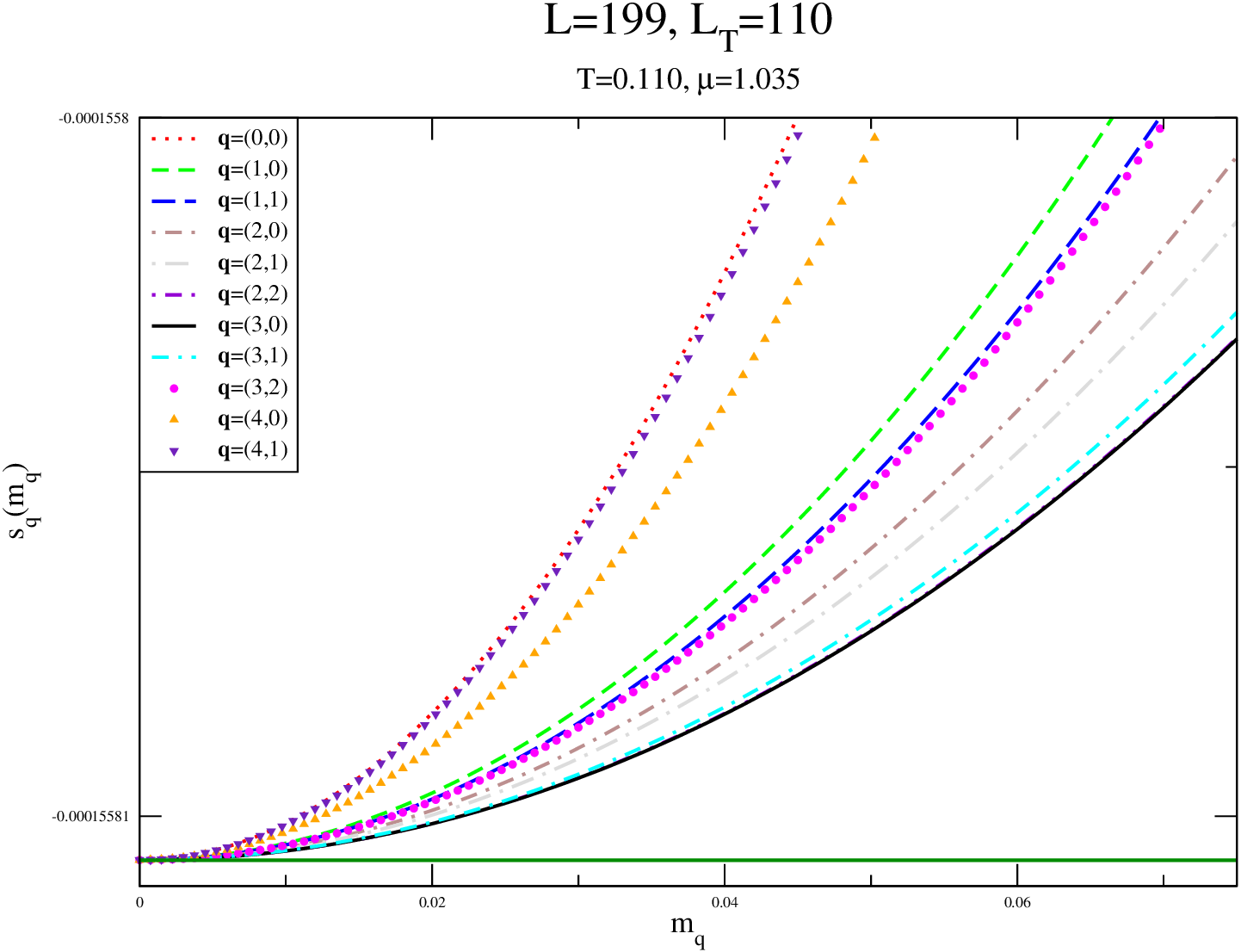}
\caption{ A sample plot of the action density $s_q(m_qm_s;\lambda(m_s),\mu m_s, L, L_T)-s_0(0;0,0,L,L)$
as a function of the condensate, $m_q$, at a chemical potential, $\mu=1.035$ and temperature
$T=0.110$ with an aspect ration of $\tau=\frac{70}{127}$. The four plots correspond to different $(L,L_T)$ with the same aspect ratio. The green horizontal line corresponds to the common value of the action density at $m_q=0$.
The y-axis label has suppressed the auxiliary dependences in  \eqn{actden} and \eqn{actdenq}
and the overall subtraction.
}
\eef{actionden3d}

Like in the two dimensional model, we plot the action density in \eqn{actden} and \eqn{actdenq} 
as a function of the scalar condensate, $m_q$, at a temperature of $T=0.110$ and
a chemical potential of $\mu=1.035$ for an aspect ratio of $\tau=\frac{70}{127}$  in \fgn{actionden3d}. Since we
are interested in finite lattice spacing effects, we have plotted $$s_q(m_qm_s;\lambda(m_s),\mu m_s,L,L_T)
- s_0(0;0,0,L,L)$$
 for all four values of $(L,L_T)$ at a same aspect ratio.
The even functions of $m_q$ have the same value at $m_q=0$ shown as a horizontal green line. Unlike the two-dimensional model
the qualitative behavior changes as one approaches the continuum limit at a fixed aspect ratio -- increasing $L$ while keeping 
$\tau=\frac{L_T}{L}$ fixed. We see that at $L=127$, six different values of ${\bf q}$, namely, $(2,0), (2,1), (2,2), (3,0), (3,1), (3,2)$,  have a minimum
at a non-zero value of $m_q$. The minimum at ${\bf q} = (3,0)$ dominates and is shown as a solid black line. 
At $L=149$, only three different values of ${\bf q}$, namely, $(2,2), (3,0), (3,1)$ have a minimum at a non-zero value of $m_q$. The
transition from ${\bf q}=(3,0)$ to ${\bf q}=(3,1)$ is barely above $\mu=1.035$ at this value of $L$ and this is evident from the minima
at these two values of ${\bf q}$ being indiscernibly close to each other. We have shown the action for ${\bf q}=(3,0)$ by a solid black line.
As $L$ is increased to $167$ and $199$, the action at all momenta shown in the plot have a trivial minimum at $m_q=0$.
We have shown the action for ${\bf q}=(3,0)$ as a solid black line all four plots. Based on the behavior at four different values of $L$
as a fixed aspect ratio, we conclude that the system is in a symmetric phase at $T=0.110$ and $\mu=1.035$ at an aspect ratio of
$\tau=\frac{70}{127}$.

\bef
\centering
\includegraphics[scale=0.325]{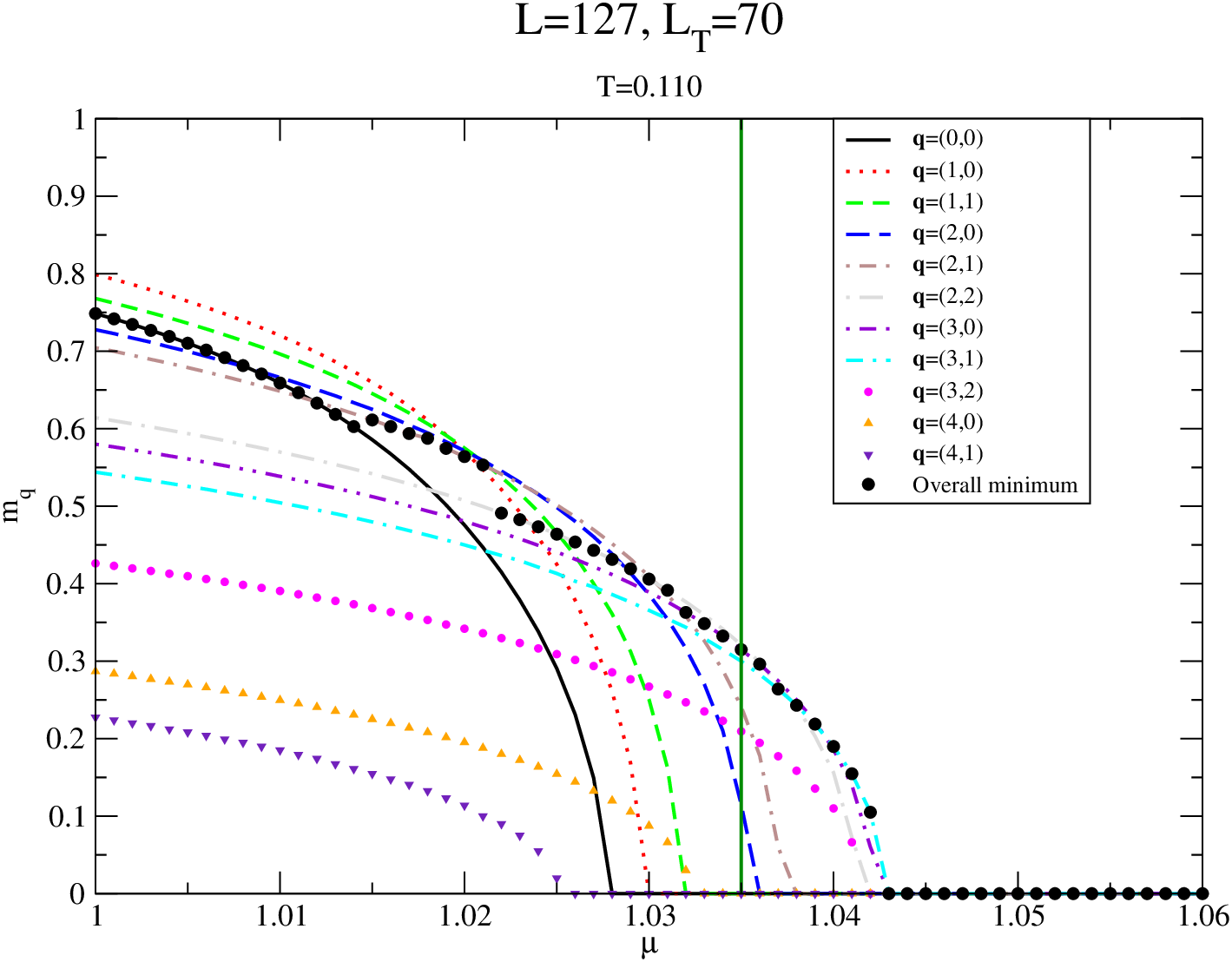}
\includegraphics[scale=0.325]{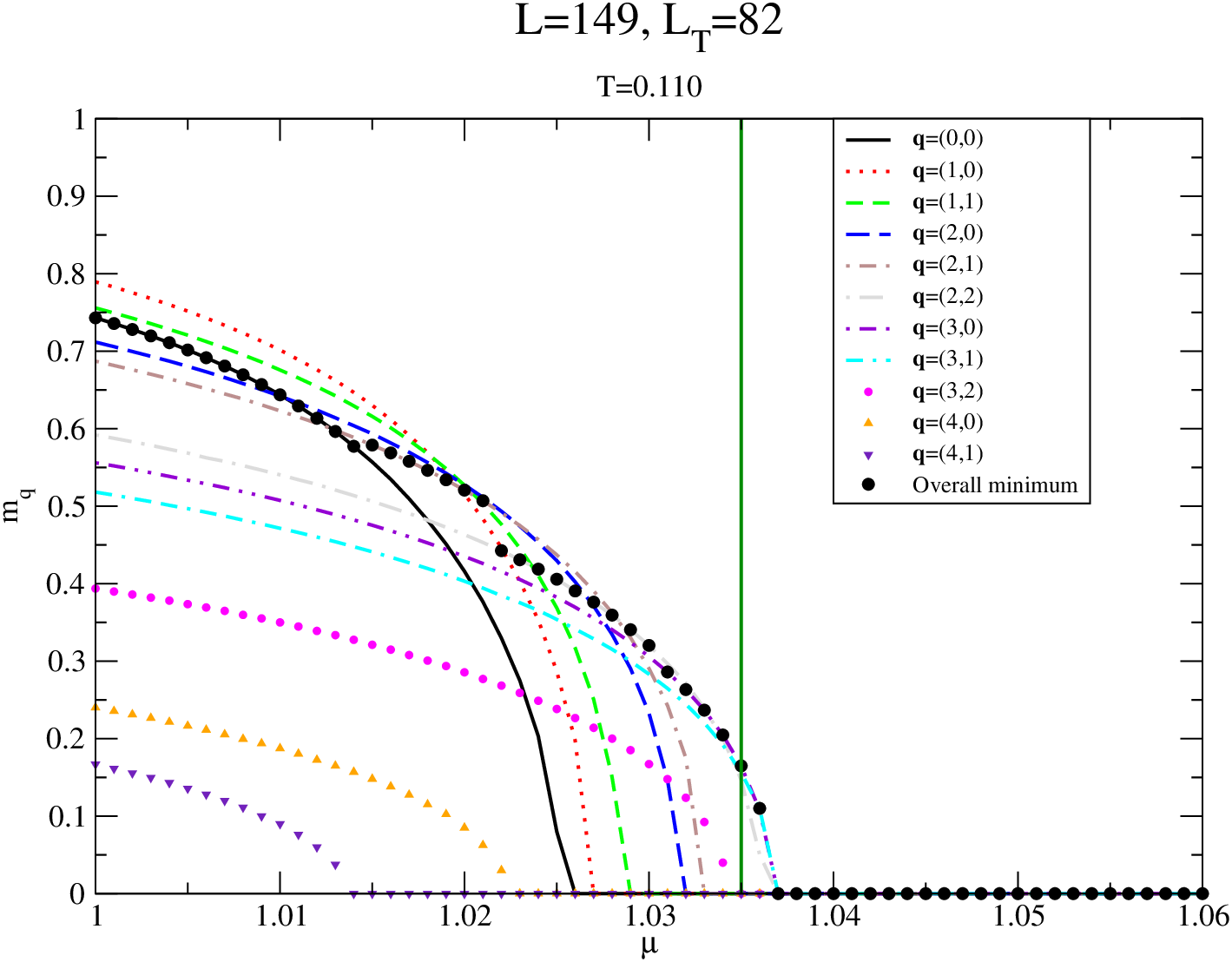}
\includegraphics[scale=0.325]{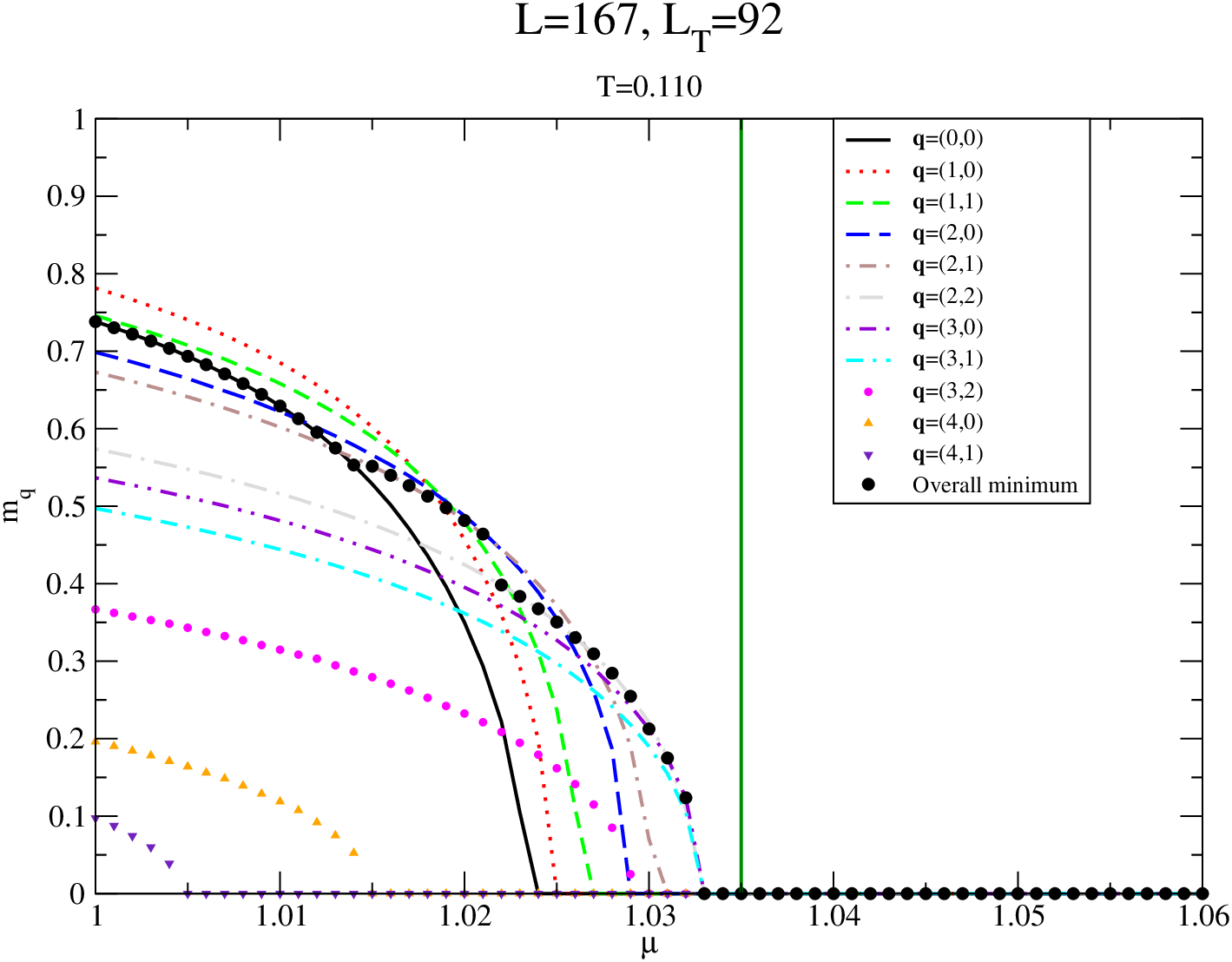}
\includegraphics[scale=0.325]{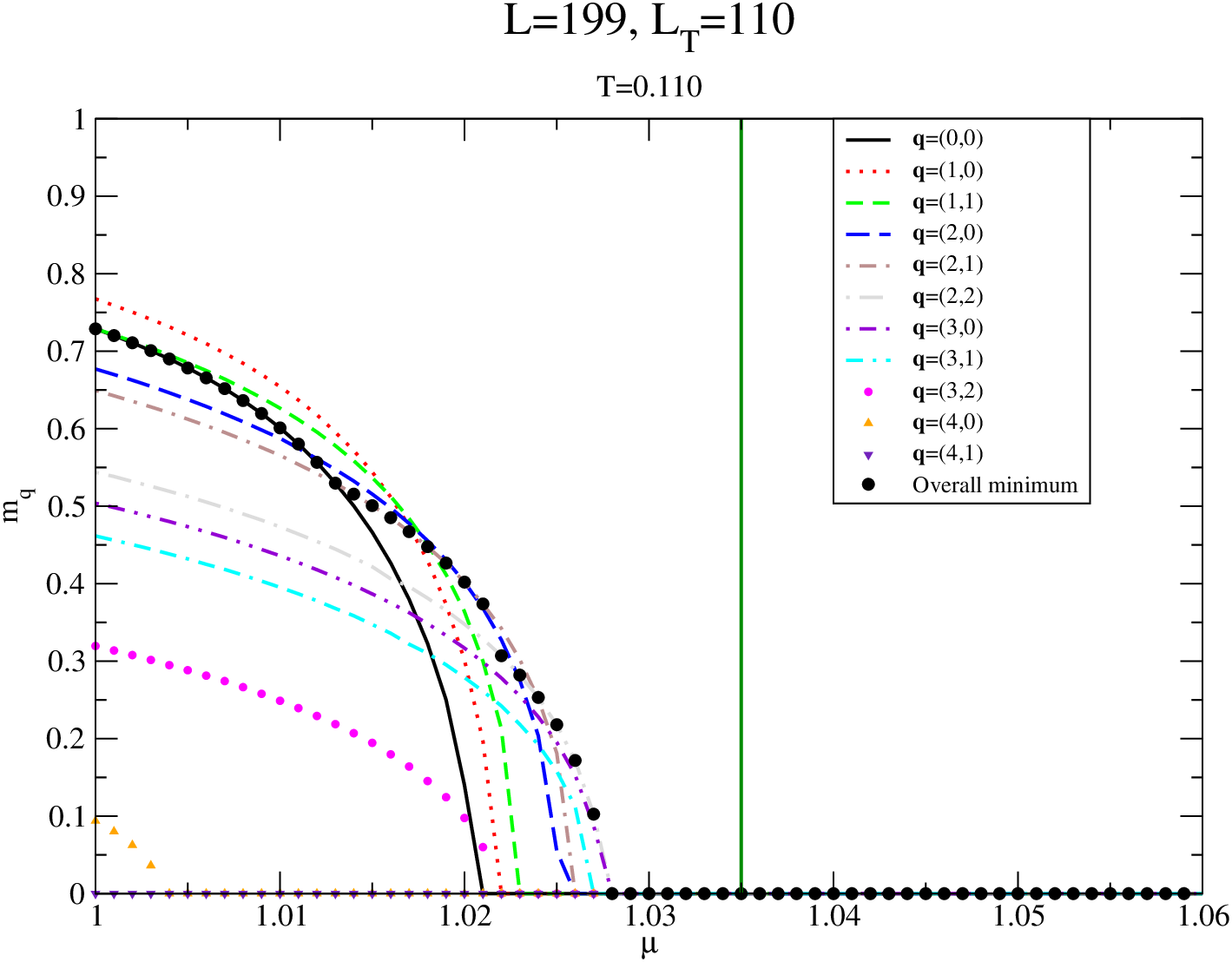}
\caption{ A sample plot of the condensate, $m_q$, as a function of the chemical potential, $\mu$, at
a temperature of $T=0.110$ with an aspect ratio of 
or $\tau=\frac{70}{127}$ . The two plots correspond to different $(L,L_T)$ with the same aspect ratio.
The green vertical line corresponds to $\mu=1.035$.
}
\eef{cond3d}

To understand the approach to the continuum limit at a fixed temperature and aspect ratio,
we plot the location of the minimum of the action density for all spatial momenta, $m_q$, for all momenta listed in \eqn{momlist3d}
as a function of the chemical potential, $\mu$, at a temperature of $T=0.110$ for an aspect ratio of $\tau=\frac{70}{127}$ in \fgn{cond3d}. Like in \fgn{actionden3d}, we have shown the results for four different values of
$(L,L_T)$ with the same aspect ratio. We have focussed on a small region of $\mu$ within the solid green box in \fgn{bigpicture3d} where we see evidence for a inhomogeneous condensate at finite lattice spacing.
The points with the black dots in \fgn{cond3d} are the value of the minima taken over all spatial momenta shown in the plot.
To facilitate comparison with \fgn{actionden2d}, we have marked $\mu=1.035$
with a green line in all four plots. The value of the inhomogeneous condensate at $L=149$ is lower that at $L=127$ and consistent with the plot in \fgn{actionden3d},
$\mu=1.035$ is very close to the transition from ${\bf q}=(3,0)$ to ${\bf q}=(3,1)$. At $L=167$ and $L=199$, the system is clearly in the symmetric phase.
We do see evidence for values of $\mu$ which favor an inhomogeneous condensate at $L=167$ and $L=199$ but the region clearly shrinks with increasing $L$.
The transitions from the homogeneous phase through various inhomogeneous phases to the symmetric phase occur as follows at the four different value of $L$:
\bea
L=127: && {\bf q}=(0,0) \to {\bf q}=(2,1) \quad(\mu=1.01487);\quad
  {\bf q}=(2,1) \to {\bf q}=(2,2) \quad(\mu=1.02138);\cr
  && {\bf q}=(2,2) \to {\bf q}=(3,0) \quad(\mu=1.03138);\quad
 {\bf q}=(3,0) \to {\bf q}=(3,1) \quad(\mu=1.03665); \cr
 &&  {\bf q}=(3,1) \to m_q=0 \quad(\mu=1.04260)\cr
L=149: &&{\bf q}=(0,0) \to {\bf q}=(2,1) \quad(\mu=1.01481);\quad
  {\bf q}=(2,1) \to {\bf q}=(2,2) \quad(\mu=1.02183);\cr
  && {\bf q}=(2,2) \to {\bf q}=(3,0) \quad(\mu=1.03055);\quad
 {\bf q}=(3,0) \to {\bf q}=(3,1) \quad(\mu=1.03501); \cr
 &&  {\bf q}=(3,1) \to m_q=0 \quad(\mu=1.03600)\cr
L=167: &&{\bf q}=(0,0) \to {\bf q}=(2,1) \quad(\mu=1.01448);\quad
  {\bf q}=(2,1) \to {\bf q}=(2,2) \quad(\mu=1.02182);\cr
  && {\bf q}=(2,2) \to {\bf q}=(3,0) \quad(\mu=1.02968);\quad
 {\bf q}=(3,0) \to m_q=0 \quad(\mu=1.03200)\cr
L=199: && {\bf q}=(0,0) \to {\bf q}=(2,1) \quad(\mu=1.01368);\quad
  {\bf q}=(2,1) \to {\bf q}=(2,2) \quad(\mu=1.02127);\cr
  && 
 {\bf q}=(2,2) \to m_q=0 \quad(\mu=1.02700)
\eea
Not only do we see evidence for the shrinking of the inhomogeneous region with a decrease in the lattice spacing, we also
see that fewer values of the wavevectors ${\bf q}$ are supported. Together, they suggest that the continuum limit does not
support a phase with a inhomogeneous condensate.

\bef
\centering
\includegraphics[scale=0.325]{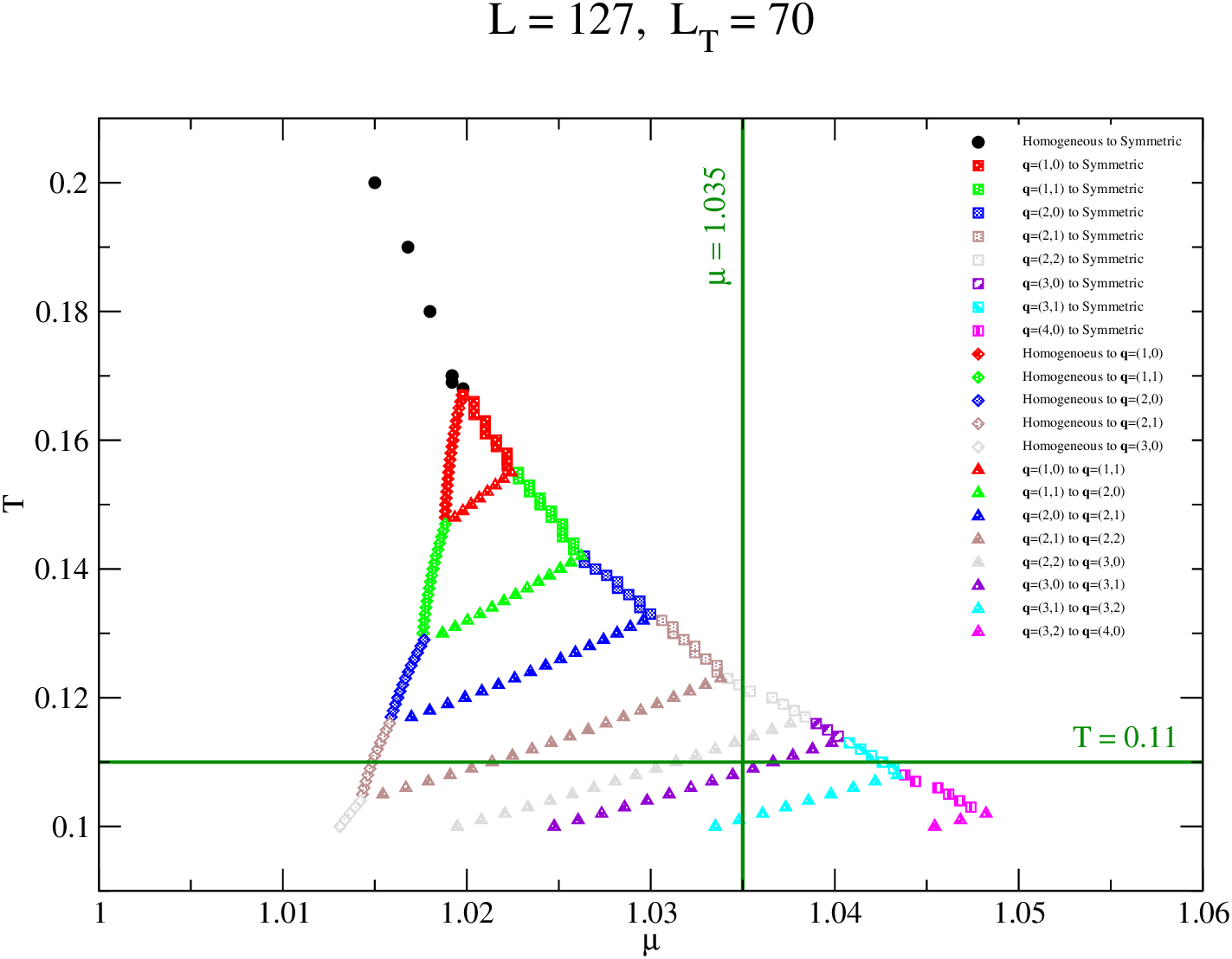}
\includegraphics[scale=0.325]{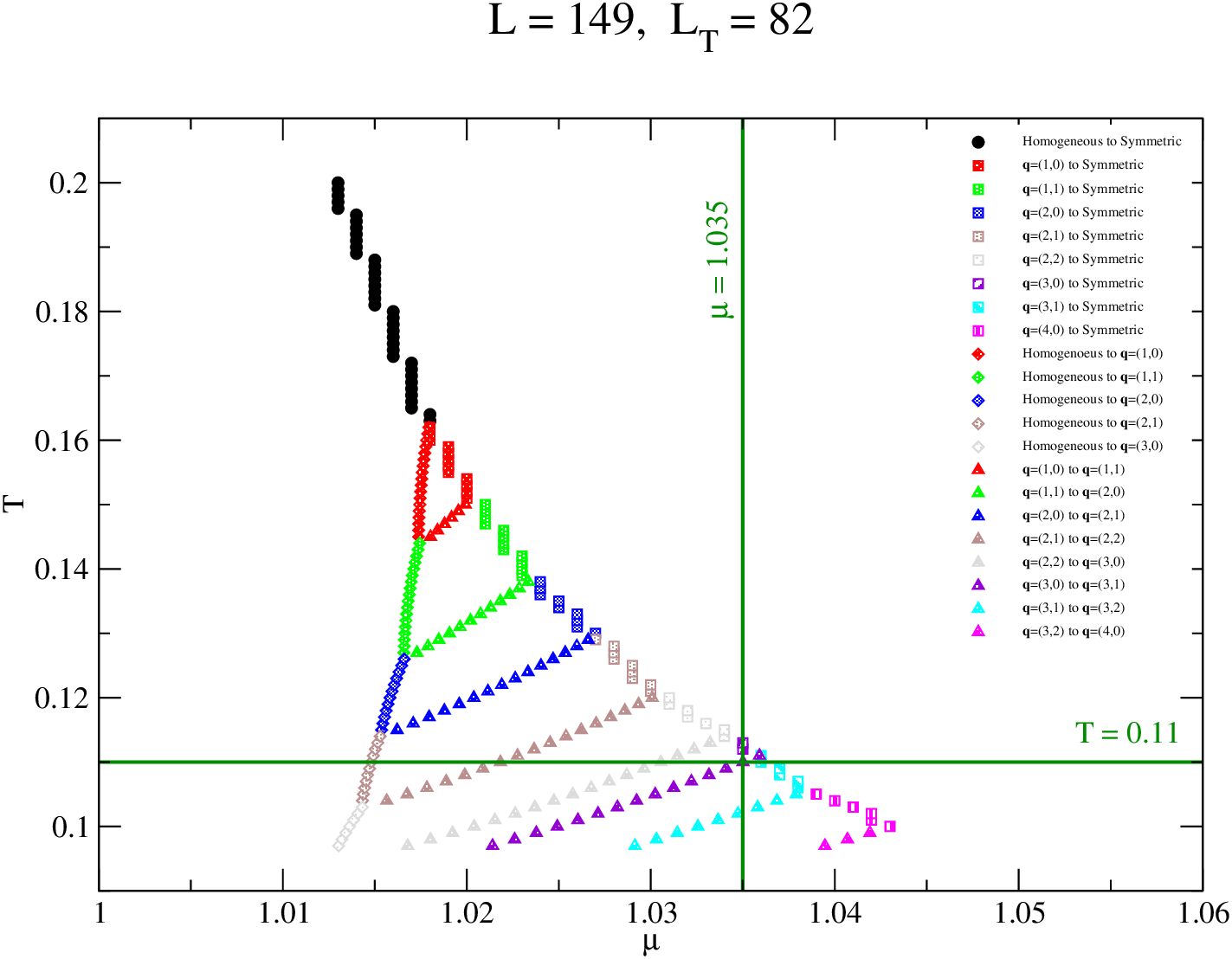}
\includegraphics[scale=0.325]{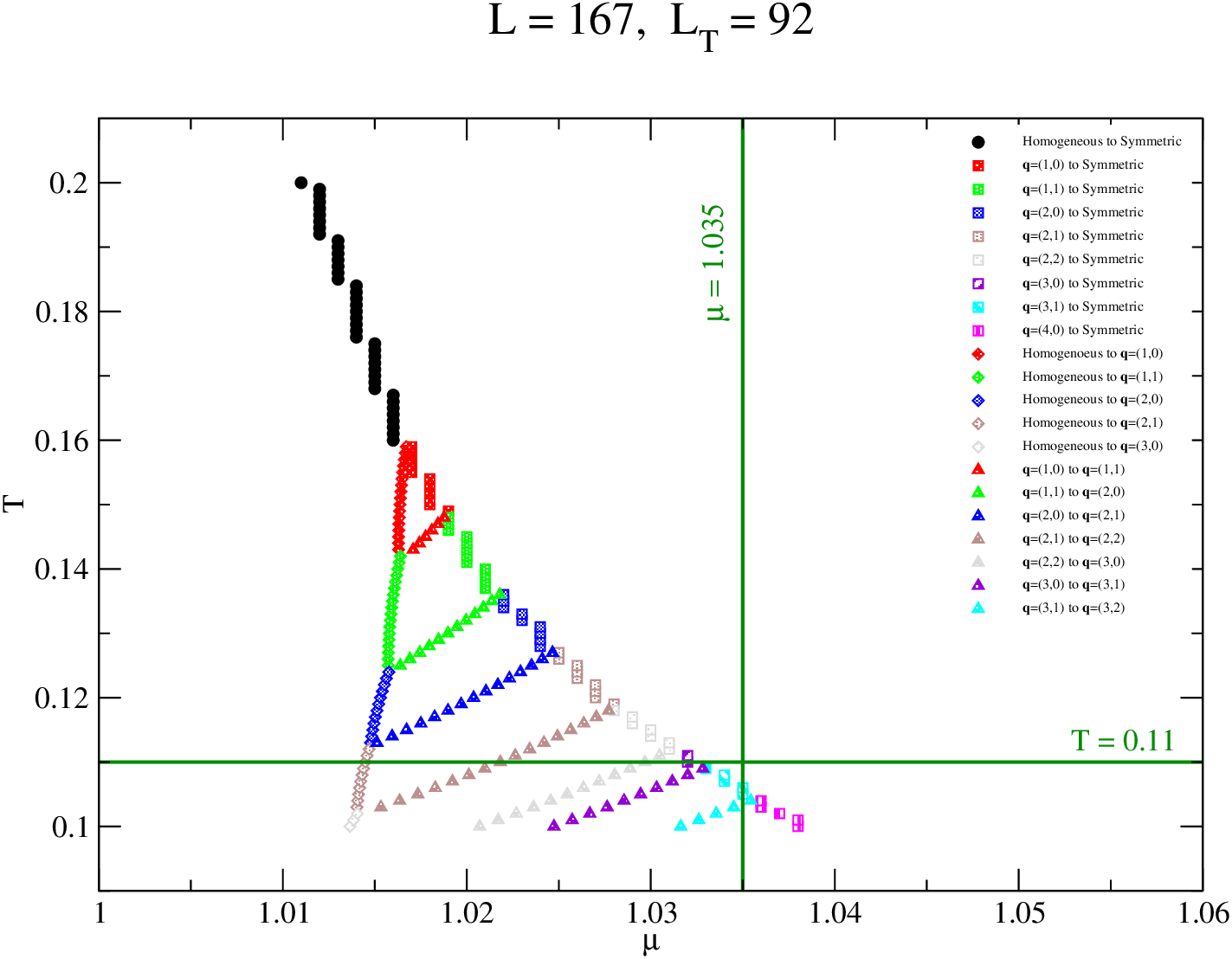}
\includegraphics[scale=0.325]{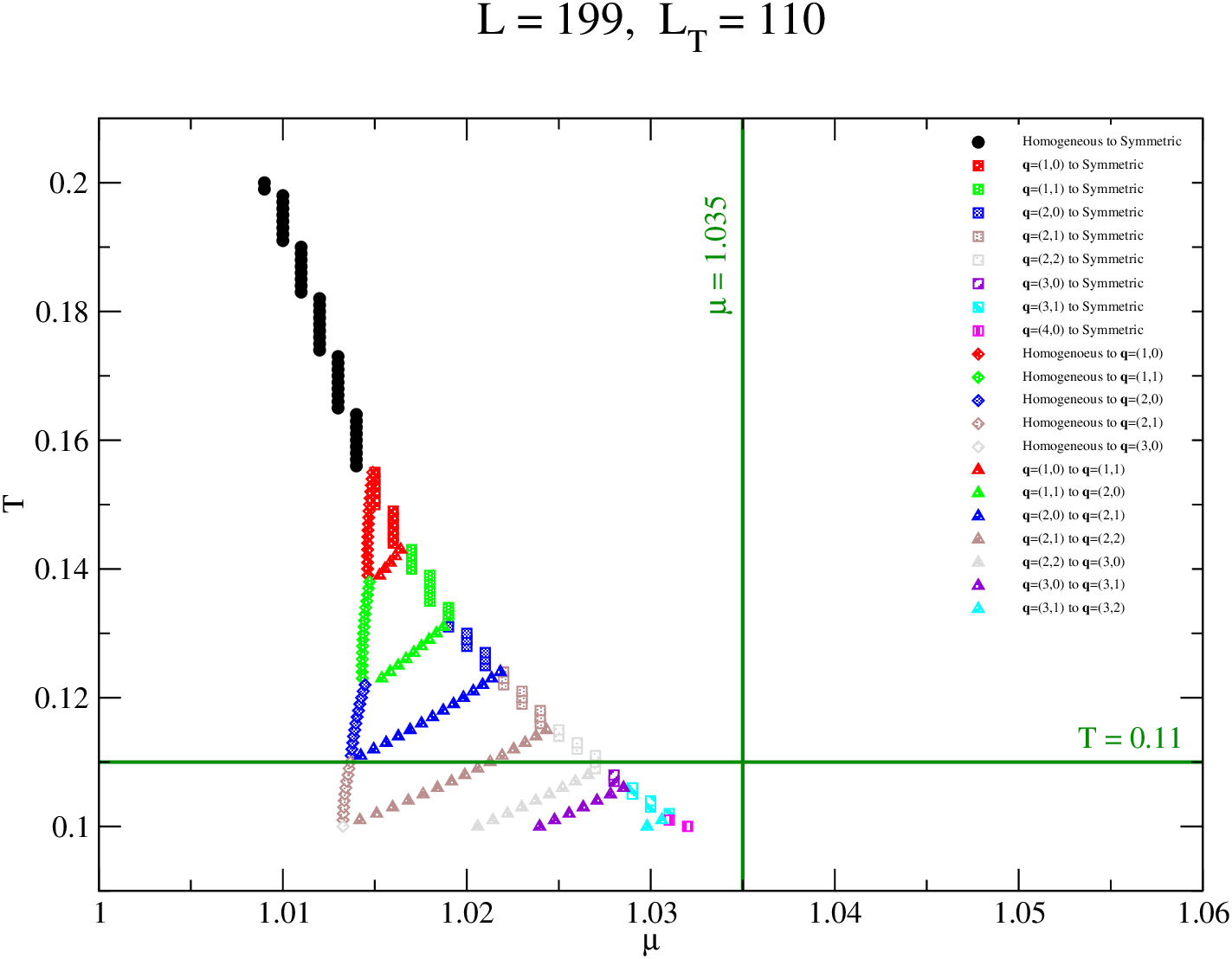}
\caption{ The full phase diagram for an aspect ratio of $\tau=\frac{70}{127}$ with four different choices of $(L,L_T)$.
}
\eef{phase3d-70}

\bef
\centering
\includegraphics[scale=0.325]{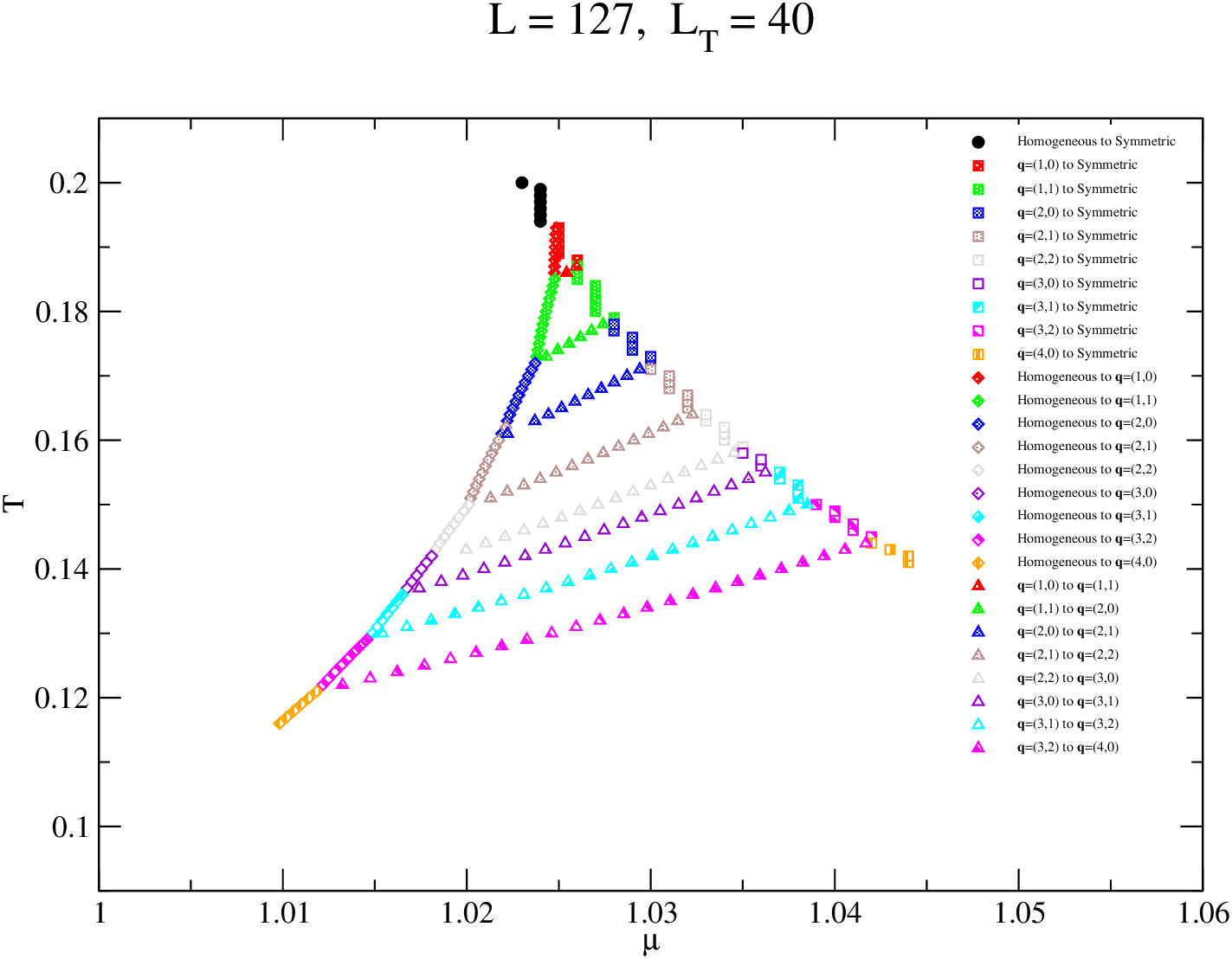}
\includegraphics[scale=0.325]{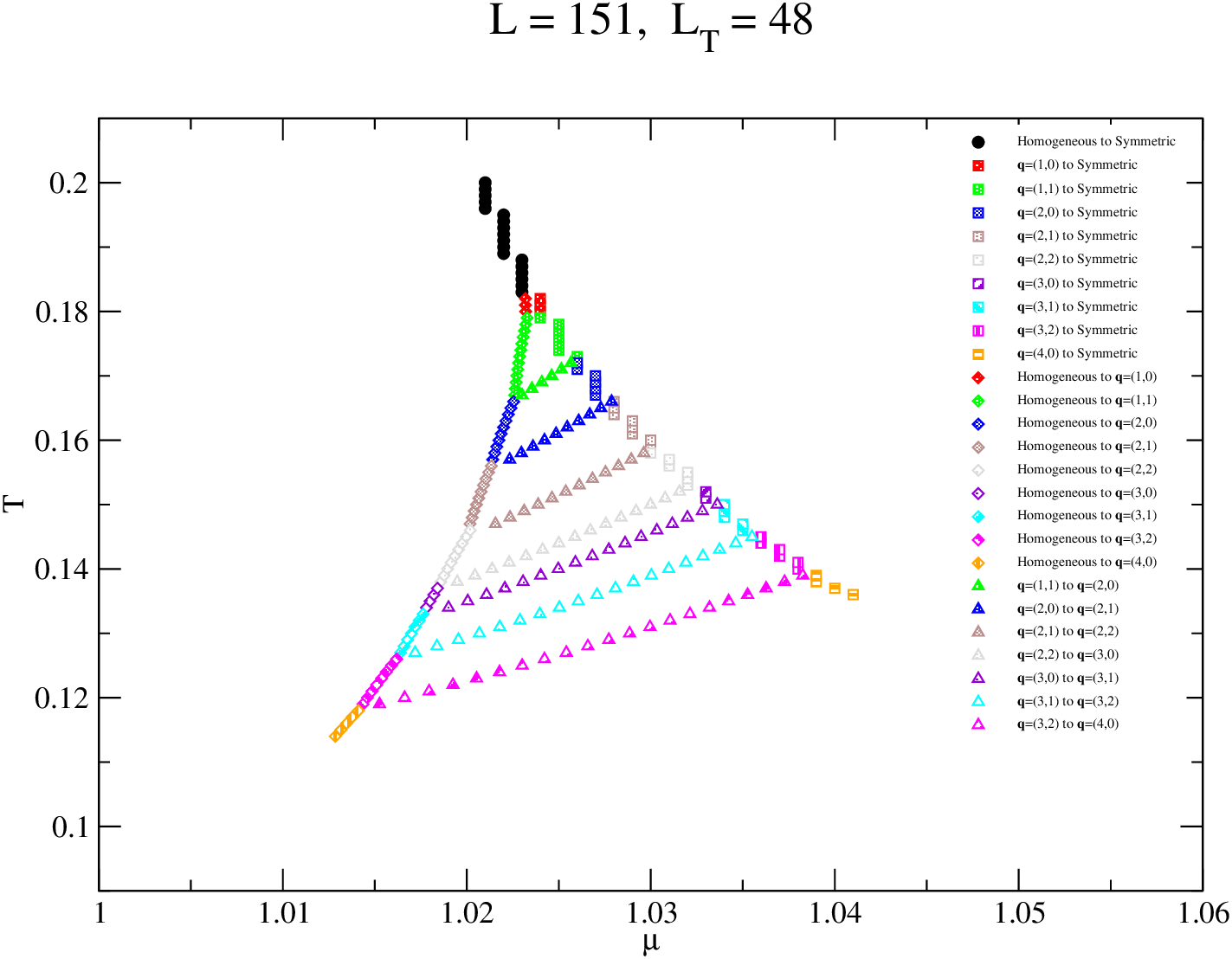}
\includegraphics[scale=0.325]{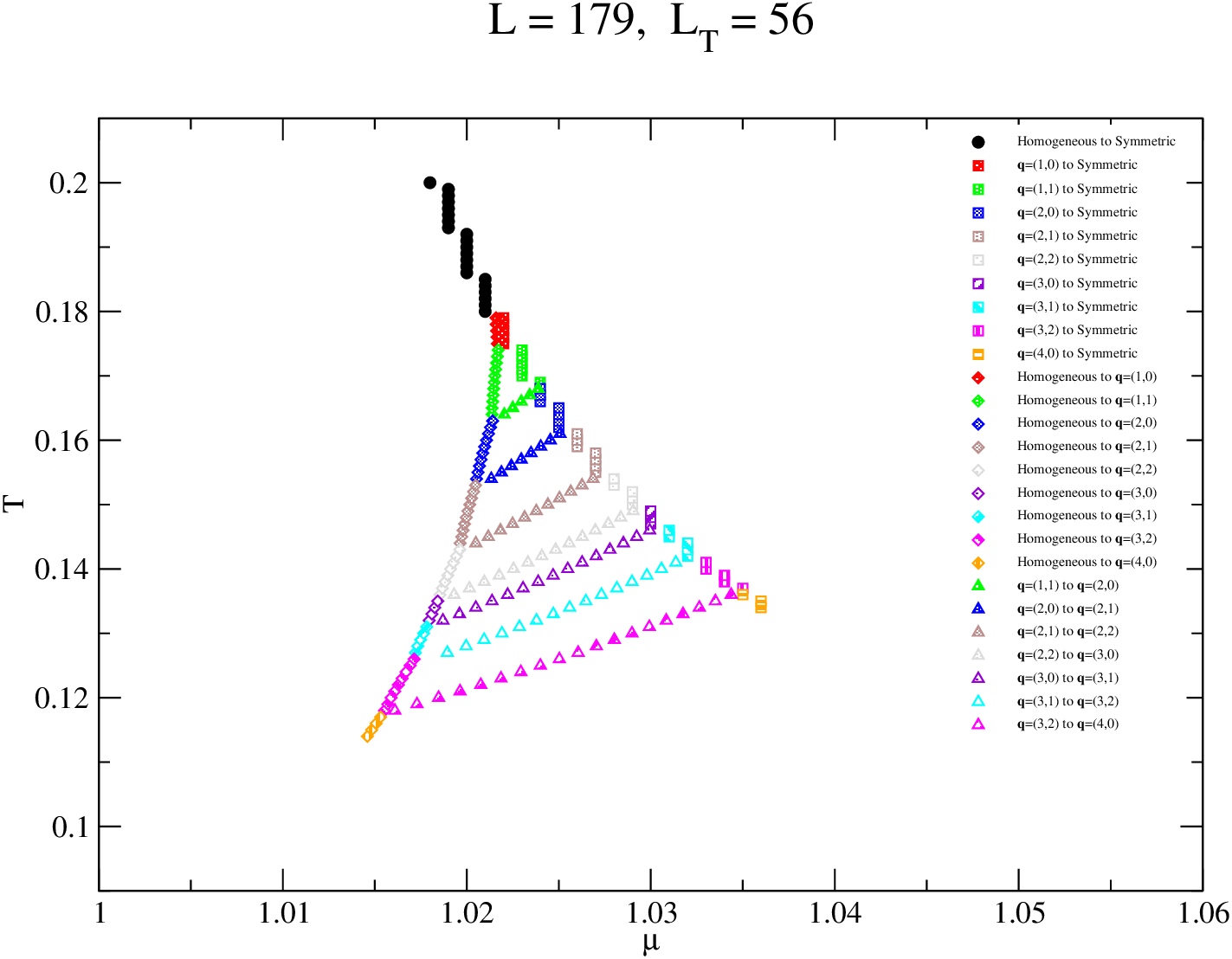}
\includegraphics[scale=0.325]{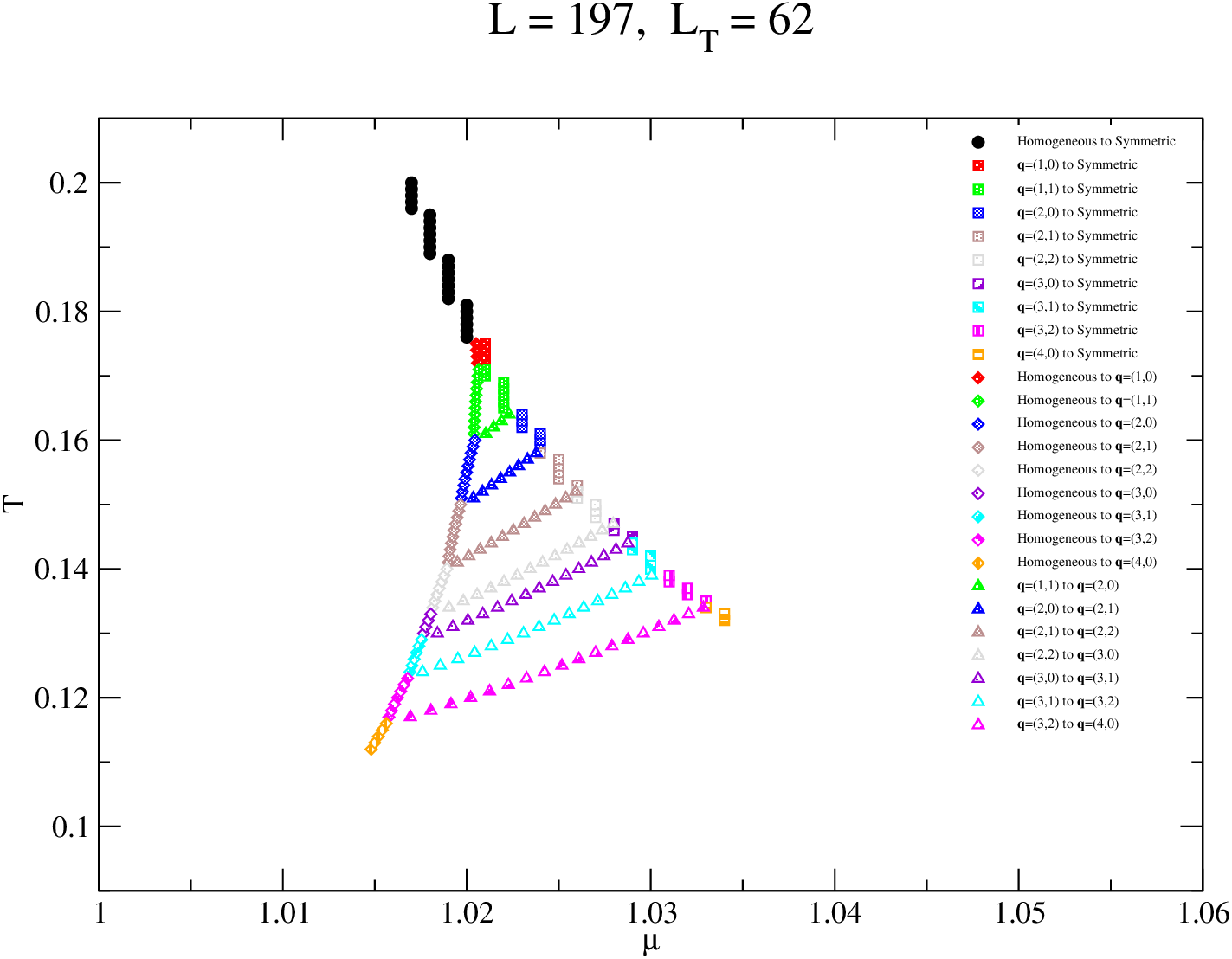}
\caption{ The full phase diagram for an aspect ratio of $\tau=\frac{40}{127}$ with four different choices of $(L,L_T)$.
}
\eef{phase3d-40}
\bef
\centering
\includegraphics[scale=0.325]{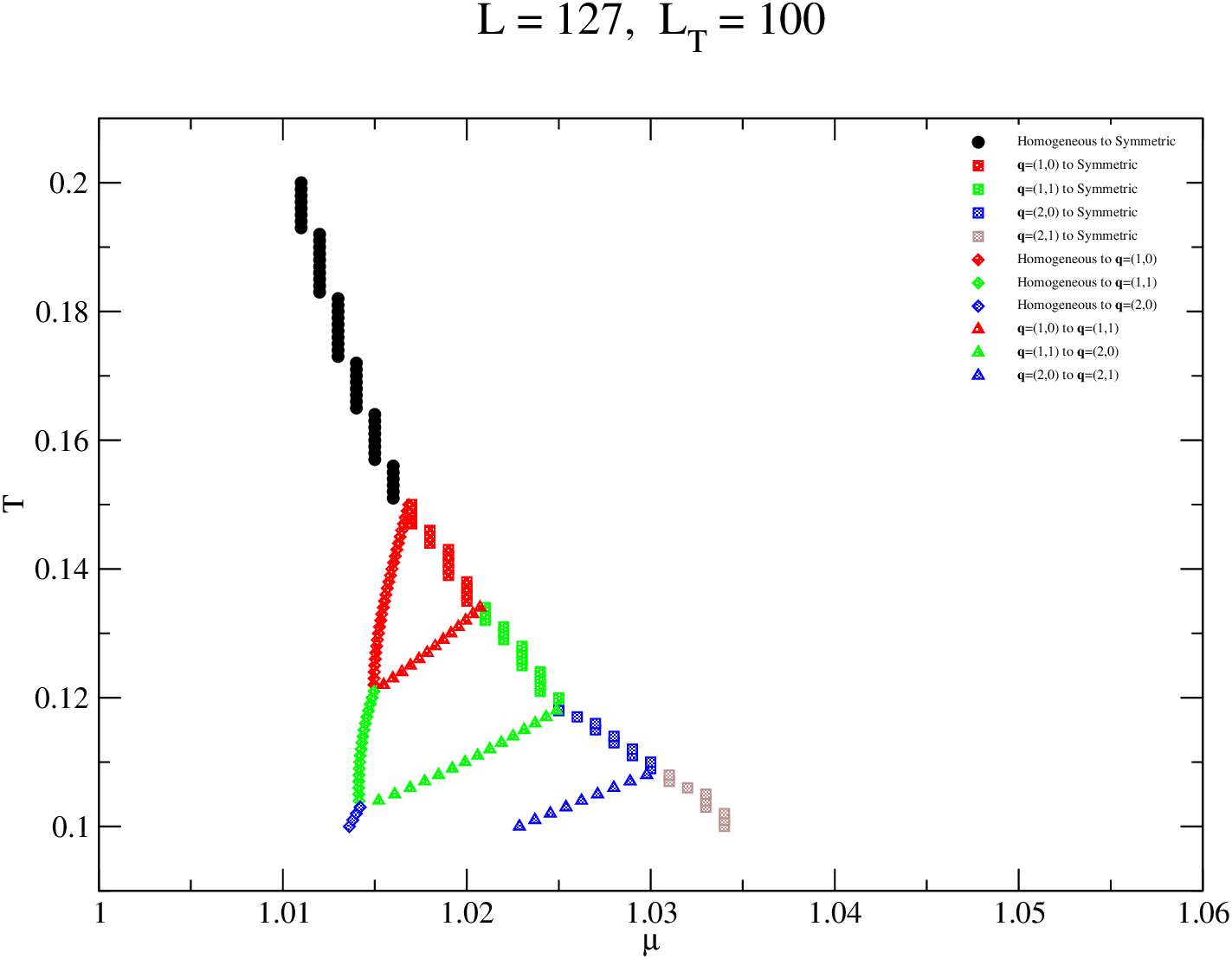}
\includegraphics[scale=0.325]{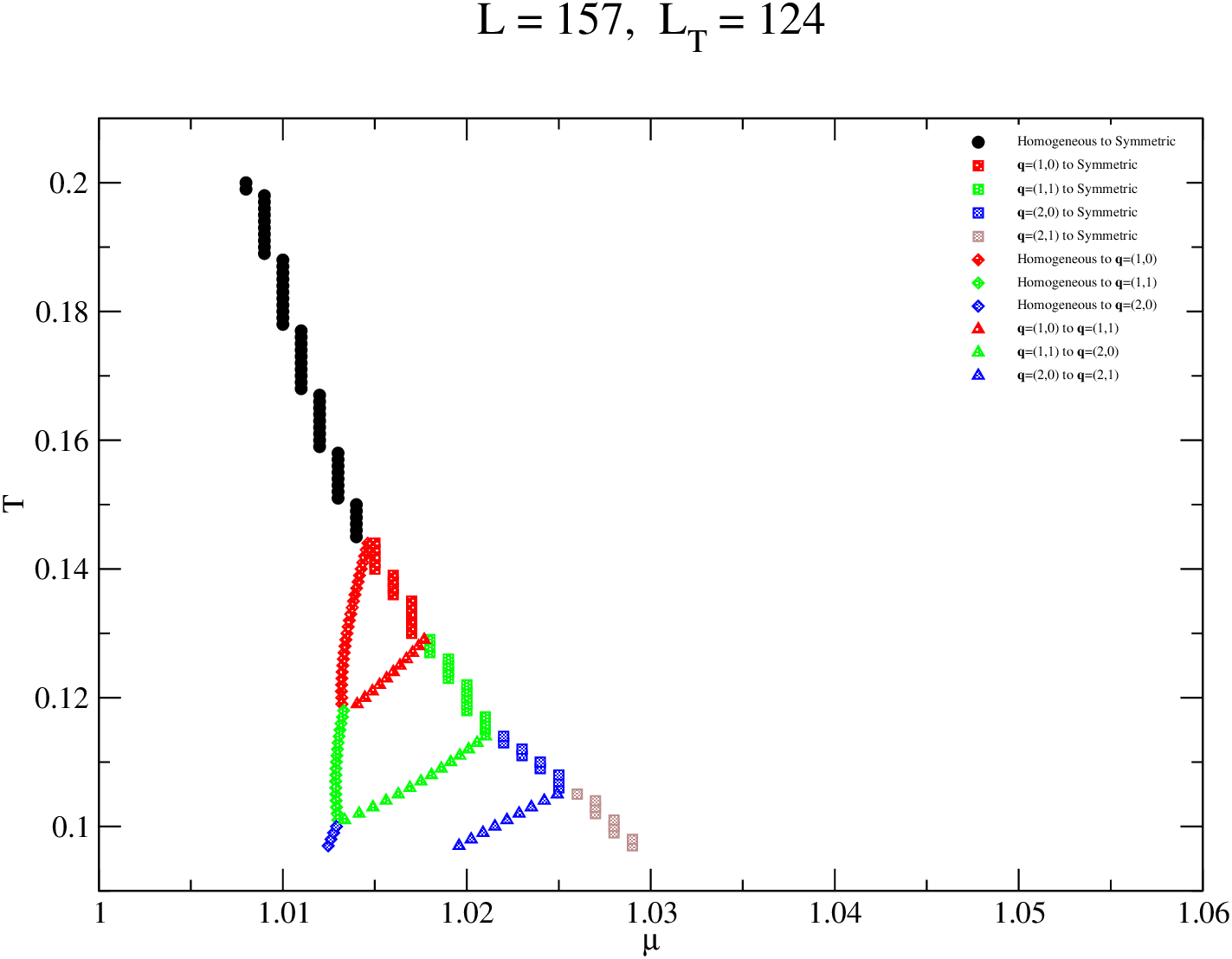}
\includegraphics[scale=0.325]{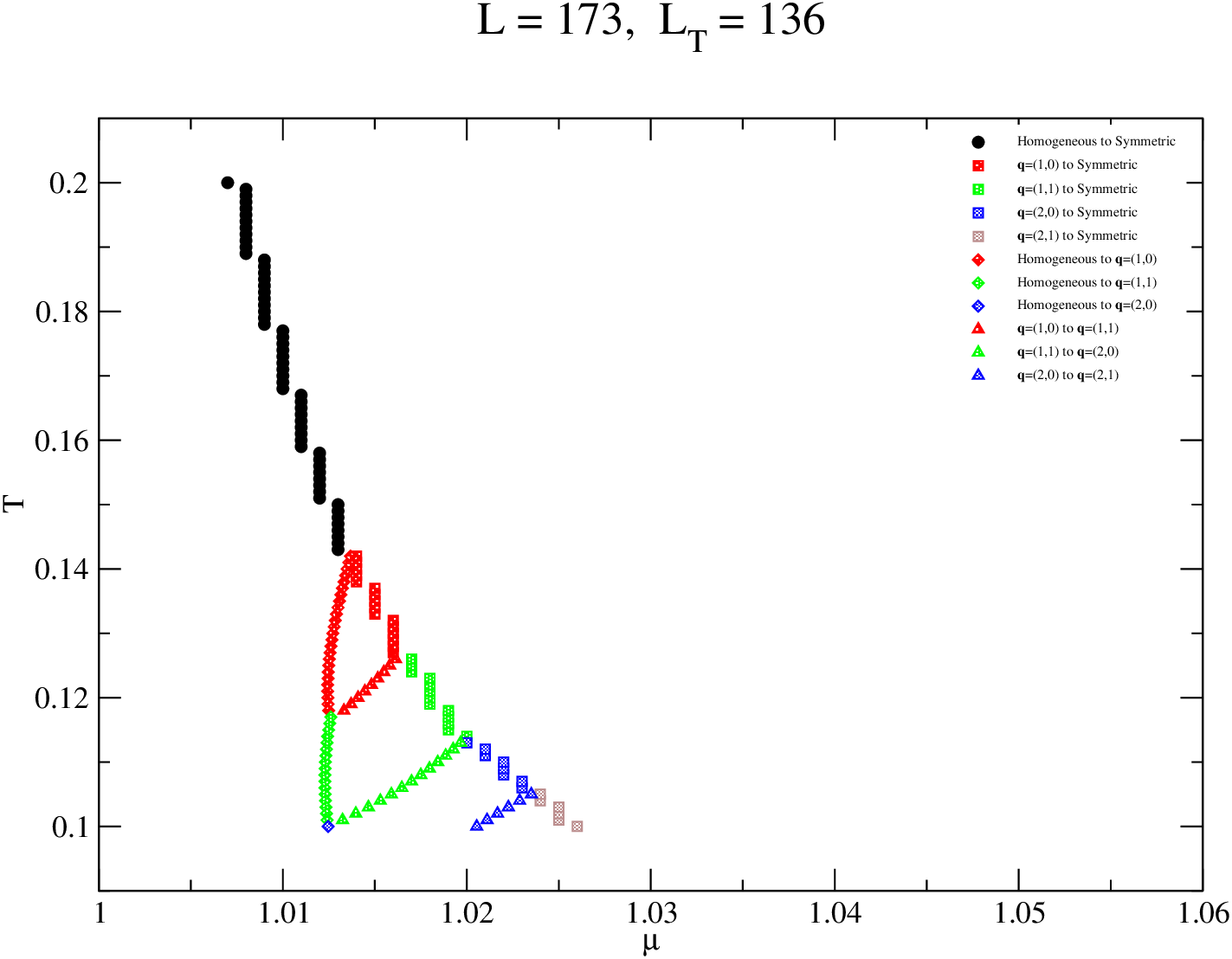}
\includegraphics[scale=0.325]{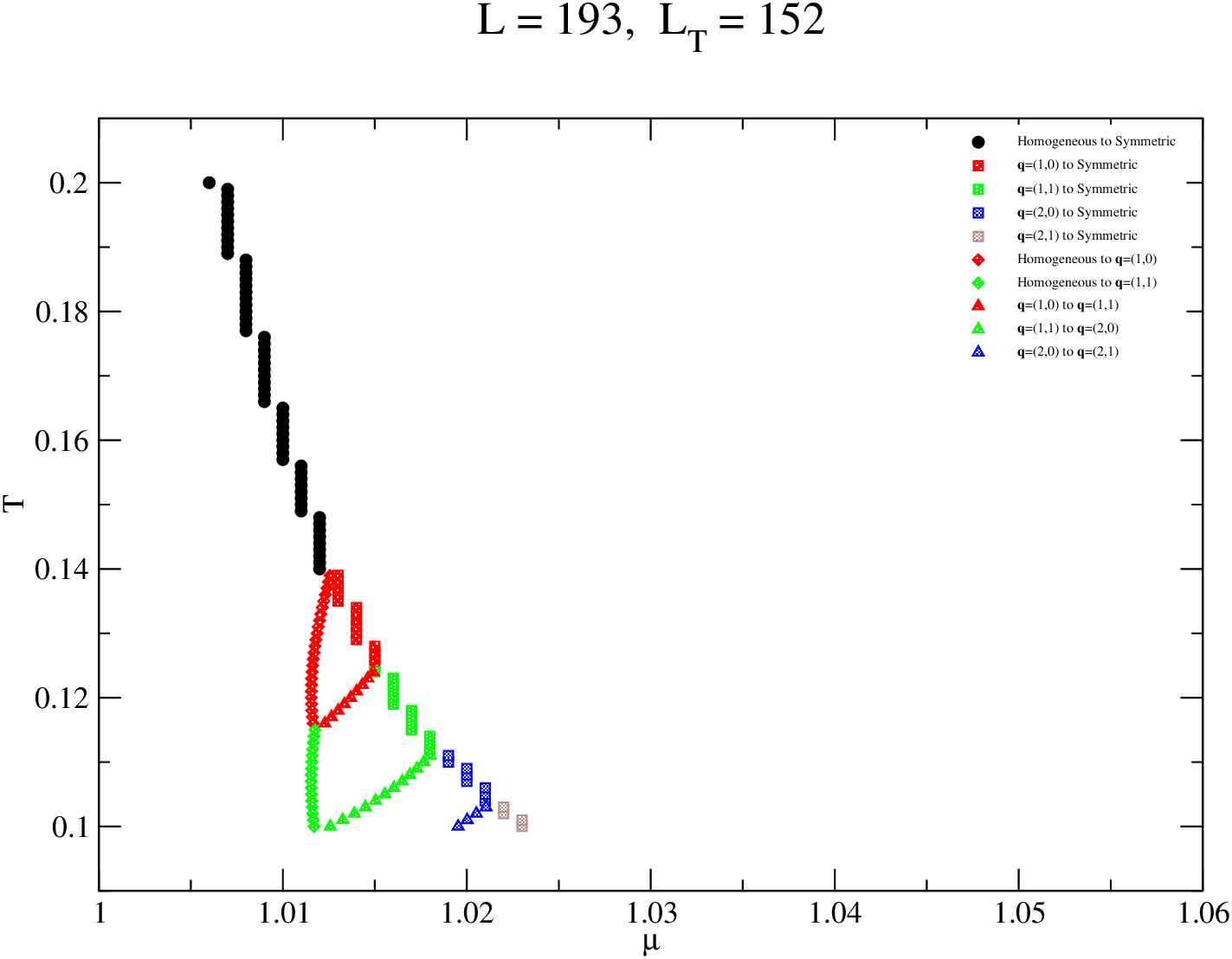}
\caption{ The full phase diagram for an aspect ratio of $\tau=\frac{100}{127}$ with four different choices of $(L,L_T)$.
}
\eef{phase3d-100}

We proceed to investigate the full phase diagram in the region enclosed in the solid green box in \fgn{bigpicture3d} for four
different lattice spacings at the aspect ratio of $\tau=\frac{70}{127}$. The four plots are shown in \fgn{phase3d-70}.
The value of $T=0.11$ and $\mu=1.035$ used in \fgn{actionden3d} and \fgn{cond3d} are marked with solid lines in all four
plots. Each plot shows all the qualitative features found for the two dimensional model in \fgn{bigpicture2d}. The crucial
difference is the clear evidence that the region with the collection of homogeneous condensates shrink in size. 
The feature discussed in \fgn{actionden3d} and \fgn{cond3d} are see in these four plots: The width of the region with an
inhomogeneous condensate shrinks at $T=0.11$ as the lattice spacing is reduced; the system is well inside the
phase with a wavevector of ${\bf q}=(3,0)$ when $(L,L_T)=(127,70)$ and $\mu=1.035$; the system is close to
the transition from ${\bf q}=(3,0)$ to ${\bf q}=(3,1)$ when $(L,L_T)=(149,82)$; and the system is well inside
the symmetric phase when $(L,L_T)=(167,92), (199,110)$ but the system can be found to be in a phase with an
inhomogeneous condensate for smaller values of $\mu$. In addition to the shrinking of the width of the phase
with an inhomogeneous condensate at a fixed $T$, we also observe that the point $(\mu_t,T_t)$ where the transitions
from (a) homogeneous phase to symmetric phase; (b) homogeneous phase to ${\bf q}=(1,0)$ inhomogeneous phase;
and (c) ${\bf q}=(1,0)$ inhomogeneous phase to symmetric phase; meet, moves to smaller temperature and chemical
potential with decreasing lattice spacing. In spite of the shrinking of the region with an inhomogeneous condensate, the
integrity of the various lines that separate inhomogeneous phases with two different wavevectors is maintained.
Like in the two dimensional model where the transition lines within the inhomogeneous phase separated phases with
wavenumbers that were different by one integer, the transition lines in the three dimensional model separates phases
with magnitude of wavevectors that are consecutive.

In order to understand the effect of non-zero aspect ratio, we compare the four phase diagrams at the aspect ratio
of $\tau=\frac{70}{127}$ with two other aspect ratios. The four plots with $\tau=\frac{40}{127}$ are shown in \fgn{phase3d-40}
and the four plots with $\tau=\frac{100}{127}$ are shown in \fgn{phase3d-100}. All three sets of plots cover the entire region
enclosed by the solid green box in \fgn{bigpicture3d} to facilitate a comparison.
When the four plots in the two sets
are compared within the set, the entire behavior is same as the one seen at $\tau=\frac{70}{127}$. When the plots are
compared across sets, we see that the width of the inhomogeneous phase as a fixed $T$, say $T=0.11$ is larger at $\tau=\frac{40}{127}$
compared to $\tau=\frac{70}{127}$ and smaller at $\tau=\frac{100}{127}$
compared to $\tau=\frac{70}{127}$. Since the values of $L_T$ at smaller $\tau$ are smaller than at larger $\tau$, the lattice spacing
is probably controlled by $L_T$ and not $L$. In a Hamiltonian analysis like the one in~\cite{Thies:2003kk}, $L_T$ is infinite and  if
one discretizes the spatial degrees of freedom appearing in the Hamiltonian, one is working with a very large aspect ratio. Our analysis
suggests that the width of the inhomogeneous region probably goes to to zero in the limit of infinite aspect ratio even at a finite lattice
spacing. If our aim is to extract the phase diagram at finite temperature and infinite spatial extent, we should approach the limit of
zero aspect ratio. Since there is clear evidence that the region with an inhomogeneous condensate shrinks for all three values of
the aspect ratio studied here, we conclude that the system is either in an phase with a homogeneous condensate or a symmetric phase
at all aspect ratios. The phase diagram shown in \fgn{bigpicture3d} is the expected result at all aspect ratios and the transition line
given by \eqn{critical-3d} remains true after our analysis.

\section{Conclusions}

We have studied the phase diagram of the Gross-Neveu model in the limit of large number of flavors in two and three dimensions.
The presence or absence of a fermion bilinear condensate can be found by solving the gap equation and obtaining the global
minimum of the effective action of the condensate. Assumption of translational invariance implies an uniform condensate and
the solution to the gap equation can be obtained by simple analytical means~\cite{Wolff:1985av,Klimenko:1987gi,Rosenstein:1988dj}. It has been
shown that there exist regions in the $(\mu,T)$ plane for the two dimensional model where a non-uniform condensate produces
a minimum that is lower than the one given by a uniform condensate~\cite{Thies:2003kk}. This suggests that one has to explore
the action density for a general form of the condensate and find a global minimum. This is a difficult problem and one attempts
to find a solution by assuming some form for the condensate and find a solution either by some pseudo-numerical method~\cite{Thies:2003kk}
or by finding an analytical solution to the gap equation~\cite{Basar:2009fg}. In order to confirm that this is indeed the global minimum,
one could use a lattice formalism where the algorithm will hopefully find the global minimum. A lattice computation suffers from
lattice spacing effects and one needs to show that the global minimum found on the lattice survives the continuum limit. This is expected
to be relatively simple if the condensate is uniform since continuum computations for this specific case show that the cut-off is trivially
removed. This is not expected to be the case when the condensate is not uniform. 

A recent lattice computation of the Gross-Neveu model in two dimensions with eight flavors has shown clear evidence for a
non-uniform condensate that has a specific momentum~\cite{Pannullo:2019prx}. This motivated us to study the effective
action of Gross-Neveu model
in the limit of large number of flavors with a condensate that has a specific momentum. In order to find the minimum of the
action, we resorted to the lattice formalism where we could use a formula for the determinant of blocked tri-diagonal matrix~\cite{2007arXiv0712.0681M}
after making some small modifications. In order to obtain the continuum limit we found it useful to study the model at a fixed
aspect ratio, $\tau=\frac{1}{\ell T}$ where $\ell$ is spatial extent. We were able to find the minimum over a wide range of
momenta and convince ourselves we have found the {\sl absolute} minimum when restricted to a condensate with a
fixed momentum. Our analysis does not consider the possibility where the condensate has more than one momentum component.

In the confines of our analysis with a condensate that has fixed momentum which can take on arbitrary values, we have shown that the
two dimensional Gross-Neveu model with infinite number of flavors has many phases at a non-zero aspect ratio. The system can
be in a phase with a condensate that has zero momentum, non-zero momentum or it can be in a symmetric phase. The transition
lines between the various phases are well defined at each aspect ratio. Some of the key features, like the critical temperature
at zero chemical potential, critical chemical potential at zero temperature, tri-critical point and the momentum as a function of the
chemical potential along the critical line that separates the inhomogeneous phase from the symmetric phase have analogs at
non-zero aspect ratio. These features consistently match with the numbers and behavior at zero aspect ratio. 
The new feature of the analysis performed here is the presence of finer details within the inhomogeneous phase -- there are
lines of transition that separate momenta with wavenumbers that differ by one. 

\bef
\centering
\includegraphics[scale=0.325]{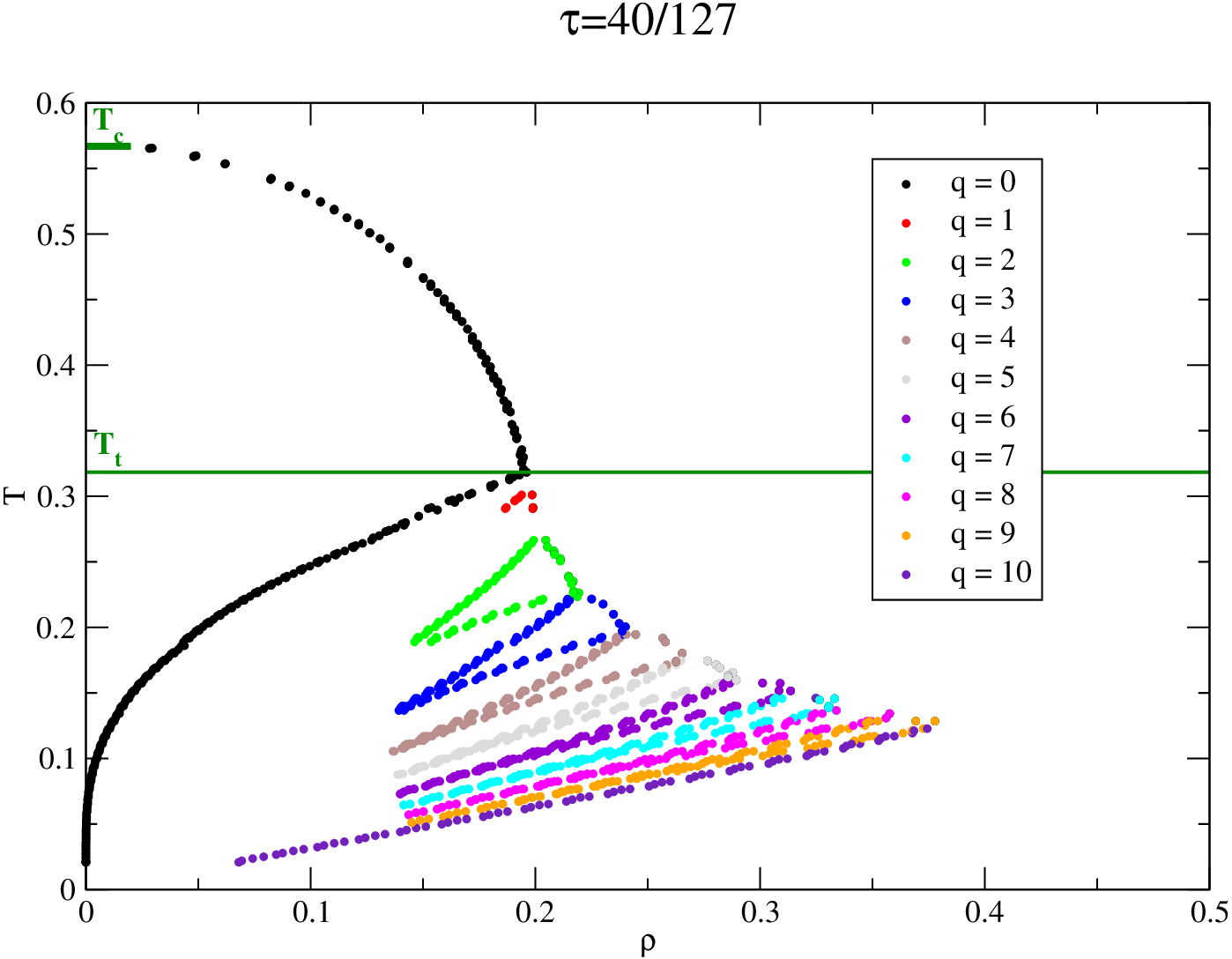}
\includegraphics[scale=0.325]{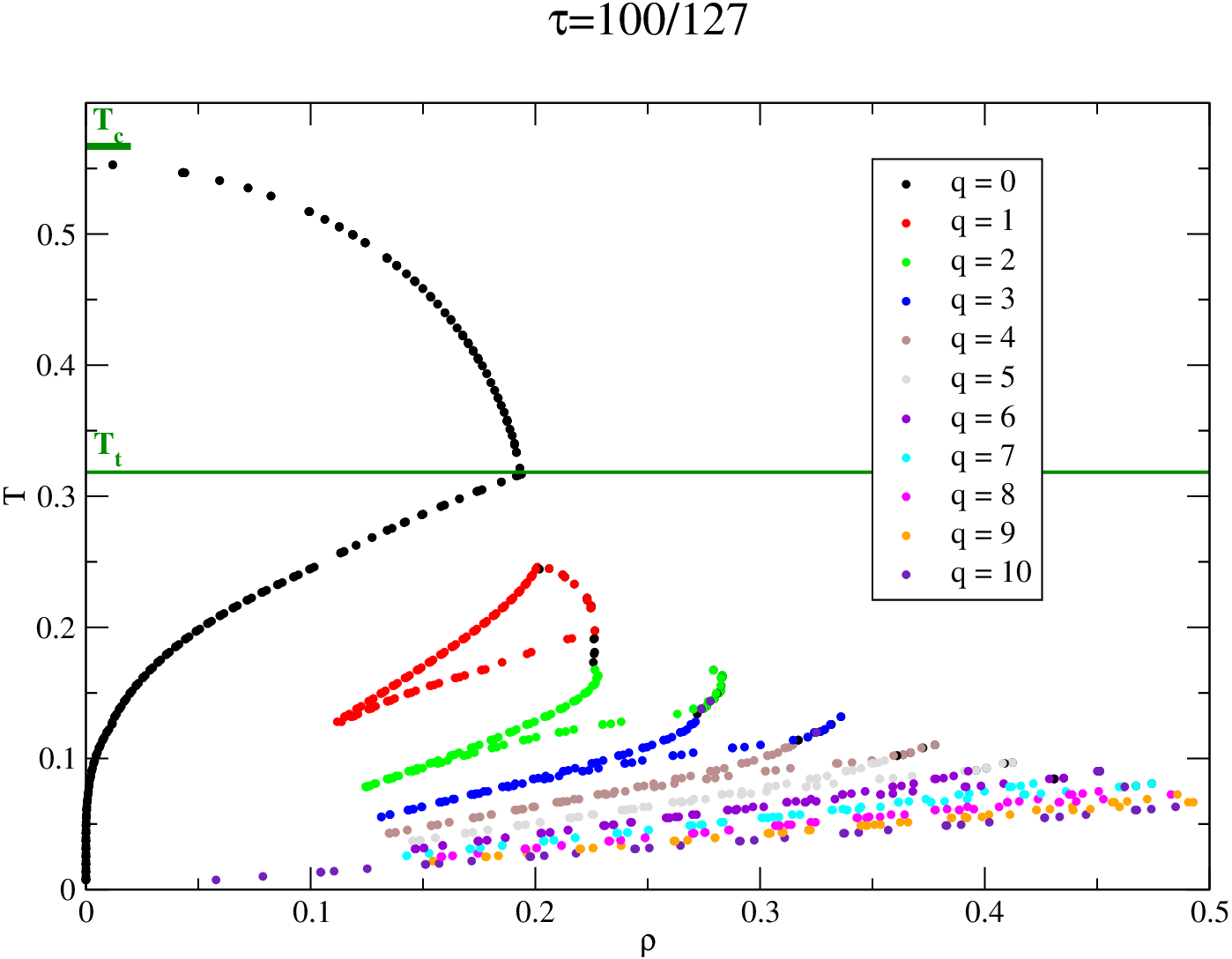}
\caption{ The phase diagram of the two dimensional model in the temperature-density plane for two different aspect ratios.
}
\eef{phase2d-trho}

In order to further understand the effects of using a condensate with a fixed momentum, we compute the number density defined as the
derivative of the action density with respect to the chemical potential. We account for a factor of four since we are using naive fermions.
Like the condensate which shows discontinuous behavior at the
transitions between the different inhomogeneous phases (as seen in \fgn{cond2d}), we also expect the number density to jump
discontinuously across the different inhomogeneous phases. For each point $(\mu, T)$ in the full phase diagram at a fixed aspect ratio,
we computed the density on either side of that point defined by the appropriate wavenumber. The resulting phase diagram is shown in \fgn{phase2d-trho}
for the smallest and the largest aspect ratio. The jumps in the number density are evident since the value for wavenumber $q$ and wavenumber $(q+1)$
at a fixed temperature are separated. Just as in the case of the phase diagrams in the $\mu-T$ plane shown in \fgn{phase-diagram-2d}, we
see the lines with a fixed $q$ coming closer to each other as the aspect ratio does down. The pinching off that one sees for each $q$
in \fgn{phase2d-trho} is a consequence of using finite number of points in the $\mu-T$ plane -- it is plausible that each of the lines extend all the way to $q=0$.
The jumps in the number density will get smaller if the lines with fixed $q$ get closer as expected for smaller aspect ratios. The boundaries of the phase diagram in the $\rho-T$ plane agrees quite well with the corresponding diagram in~\cite{Thies:2003kk}. In addition, the location of the tri-critical temperature, $T_t$, is easier to identify.

We also performed an analysis of the three dimensional Gross-Neveu model in the limit of large number of flavors where
previous continuum analysis does not support a inhomogeneous phase~\cite{Urlichs:2007zz} but a previous lattice analysis provides evidence
for an inhomogeneous phase~\cite{Winstel:2019zfn}. Again we assumed an inhomogeneous condensate, if one exists, has a definite momentum in the spatial directions and we also studied the model at a fixed aspect ratio. We did find a phase diagram qualitatively similar to the one
in the two dimensional model at a finite lattice spacing lending support to the results in~\cite{Winstel:2019zfn}. Unlike the two dimensional
model, the region that supports the inhomogeneous phase in the three dimensional model at a finite lattice spacing shrunk as the lattice
spacing was decreased suggesting that such a phase does not exist in the continuum limit lending support to~\cite{Urlichs:2007zz}.
At the outset, the two and three dimensional models are different: An analysis assuming a homogeneous condensate has a tricitical point
in two dimensions 
along the line of transition that separates the two phases but the transition in three dimensions in entirely second order. 
Our conclusion based on our analysis restricted to condensate with a fixed momentum is that the phase diagram of the three dimensional
model only has a homogeneous phase and a symmetric phase at all aspect ratios. 

It is possible to perform an analysis on the lattice for the two dimensional model where one assumes that the condensate has two different momenta. A naive guess would be that only one survives well inside one of the inhomogeneous phases found here. On the other hand
it might reveal further structure to the transition between two inhomogeneous phases found here. It would be interesting to see if such
a finer structure survives the continuum limit.
 
\begin{acknowledgments}
The author would like to thank Philippe de Forcrand for critical comments on the first version of this preprint.
The author would like to thank Nikhil Karthik, Michael Thies, Marc Wagner and Andreas Wipf for discussions.
  The author acknowledges partial support by the NSF under grant
numbers PHY-1515446 and PHY-1913010.  
\end{acknowledgments}

\appendix

\section{Fermion determinant with a non-zero momentum condensate}\label{sec:detqform}

The formula for the fermion determinant in \eqn{fermat} is~\cite{2007arXiv0712.0681M}
\be
\det D_c = \left(\frac{m^L_q}{2}\right) ^{2L_1} \det (T - 1);\qquad T = T_{L-1} T_{L-2} \cdots T_1 T_0;\qquad T_i = \begin{pmatrix} -\frac{2}{m^L_q} D_i & -1 \cr 1 & 0 \end{pmatrix}.
\ee 
We have suppressed the dependence of $D_c$ on the momenta and chemical potential since the details below apply to both two
and three dimensions.
Even though, we will only use this formula when $L$ is prime, we will work out the details below for an arbitrary $L$.
As it stands the factorization will not work numerically well close to $m^L_q=0$. To resolve this issue, we write
\be
T_i = \begin{pmatrix} -\frac{2}{m^L_q} & 0 \cr 0 & 1 \end{pmatrix} \bar T_i
\begin{pmatrix} 1 & 0 \cr 0 & -\frac{2}{m^L_q} \end{pmatrix};\qquad \bar T_i = \begin{pmatrix} D_i & -\frac{1}{4} {m^L_q}^2 \cr 1 & 0 \end{pmatrix}.
\ee
and obtain
\be
\det D_c = 
 \left(\frac{2}{m^L_q}\right) ^{2L} \det \left [ \bar T - \left(-\frac{m^L_q}{2}\right)^{L} \right].\label{ferdeti}
\ee
Since $\bar T$ is a $4\times 4$ matrix with
\be
\det \bar T =  \left (  \frac{m^L_q}{2} \right)^{4L_1}.
\ee
we can write
\be
\det D_c = - \left[ 1-  \left (  \frac{m^L_q}{2} \right)^{2L}\right]^2 - \left ( - \frac{m^L_q}{2} \right)^{L} \Tr \bar T -  \left ( - \frac{m^L_q}{2} \right)^{3L} \Tr \frac{1}{\bar T}
+ \frac{1}{2} \left [ \det (\bar T -1) + \det (\bar T+1) \right] .
\ee
We define
\be
\tilde T_i = \begin{pmatrix} 0 & \frac{1}{4} {m^L_q}^2 \cr -1 &  D_i \end{pmatrix}
\ee
and note that
\be
\tilde T_i \bar T_i = \bar T_i \tilde T_i = \frac{1}{4}{m^L_q}^2 ;\qquad \tilde T_i = U \bar T_i U;\qquad U = \begin{pmatrix} 0 & \frac{m^L_q}{2}\cr
\frac{2}{m^L_q} & 0 \end{pmatrix};\qquad U^2=1.
\ee
We arrive at the final expression,
\be
\det D_c = - \left[ 1-  \left (  \frac{m^L_q}{2} \right)^{2L_1}\right]^2 - \left ( - \frac{m^L_q}{2} \right)^{L_1} \Tr \left( \bar T +  \tilde T \right)
+ \frac{1}{2} \left [ \det (\bar T -1) + \det (\bar T+1) \right],
\ee
which only involves positive powers of $m^L_q$.

Due to the chiral nature of $D_i$ in two dimensions, it is possible to block diagonalize $\bar T_i$ by
\be
\bar T_i = P^t \begin{pmatrix} 0 & X_i \cr Y_i & 0 \end{pmatrix} P;\qquad P=\begin{pmatrix} 1 & 0 & 0 & 0 \cr 0 & 0 & 0 & 1\cr 0 & 1 & 0 & 0 \cr 0 & 0 & 1 & 0
\end{pmatrix};\qquad P^t P =1,
\ee 
where
\be
X_i = \begin{pmatrix} c(k^i,k_2,\mu_L)  & -\frac{{m^L_q}^2}{4} \cr 1 & 0 \end{pmatrix};\qquad Y_i = \begin{pmatrix}  -c^*(k^i,k_2,-\mu_L) & -\frac{{m^L_q}^2}{4} \cr 1 & 0 \end{pmatrix}.
\ee
Starting from \eqn{ferdeti}, we arrive at 
\be
\det D_c = \begin{cases} \left [ 2\left( \frac{m^L_q}{2} \right) ^{L} - \Tr \Psi \right]
\left [ 2\left( \frac{m^L_q}{2} \right) ^{L} - \Tr \Phi \right] & {\rm even}\  L;\cr
\left [ 2\left( \frac{m^L_q}{2} \right) ^{2L} - \Tr (\Psi\Phi) \right]& {\rm odd}\  L
\end{cases},
\ee
as the final expression in two dimensions
where
\bea
\Psi &=& X_{L-1} Y_{L-2} X_{L-3} Y_{L-4} \cdots X_3 Y_2 X_1 Y_0;\cr
\Phi  &=& Y_{L-1} X_{L-2} Y_{L-3} X_{L-4} \cdots Y_3 X_2 Y_1 X_0.
\eea
when $L$ is even
and
\bea
\Psi  &=& Y_{L-1} X_{L-2} Y_{L-3} X_{L-4} \cdots Y_2 X_1 Y_0;\cr
\Phi &=& X_{L-1} Y_{L-2} X_{L-3} Y_{L-4} \cdots X_2 Y_1 X_0 .
\eea
when $L$ is odd.

\bibliography{../../mynotes/biblio}
\end{document}